\documentclass[sigconf]{acmart}
\usepackage{soul}
\usepackage{hyperref}
\usepackage{enumitem}
\usepackage{listings}
\usepackage{cleveref}
\usepackage{booktabs} 
\usepackage{xspace}
\usepackage{xcolor}
\usepackage{microtype}
\usepackage{subcaption}
\usepackage{balance}
\usepackage{mathtools}
\usepackage{scalerel}
\usepackage[makeroom]{cancel}

\usepackage{tcolorbox}
\usepackage{microtype}
\usepackage[section]{placeins}
\usepackage{graphicx}
\usepackage[T1]{fontenc}
\usepackage{svg}
\usepackage{amsmath}
\usepackage{amsmath}
\usepackage[mathletters]{ucs}
\usepackage[T1]{fontenc}
\usepackage[utf8x]{inputenc}
\usepackage{pict2e}
\usepackage{wasysym}
\usepackage[english]{babel}
\usepackage{tikz}
\usepackage{array}

\usepackage{algpseudocode}
\usepackage{tablefootnote}
\usepackage[normalem]{ulem}

\DeclareMathSymbol{:}{\mathord}{operators}{"3A}

\usepackage{multirow}
 \usepackage{balance}

\def\HS{\hspace{\fontdimen2\font}}
\definecolor{darkgreen}{rgb}{0.15,0.55,0.15}
\definecolor{darkblue}{rgb}{0.1,0.1,0.5}
\definecolor{blue}{rgb}{0.01,0.40,.8}
\definecolor{darkgreen}{rgb}{0.15,0.55,0.15}
\definecolor{mred}{rgb}{.80,.12,.30}
\definecolor{grey}{rgb}{0.5,0.5,0.5}
\definecolor{Purple}{rgb}{.75,0,.85}
\definecolor{light-gray}{gray}{0.95}
\definecolor{mid-gray}{gray}{0.85}
\definecolor{darkred}{rgb}{0.7,0.25,0.25}
\definecolor{rose}{rgb}{1.0, 0.01, 0.24}
\newcommand{\red}[1]{\textcolor{red}{#1}}
\newcommand{\gray}[1]{\textcolor{grey}{#1}}

\newcommand{\blue}[1]{\textcolor{blue}{#1}}
\usepackage{algorithm}

\newtcbox{\redbox}{on line,
  colframe=white,colback=red!10!white,
  height=1em,valign=bottom,
  boxrule=0.5pt,arc=2pt,boxsep=0pt,left=2pt,right=2pt,top=1pt,bottom=1pt}
\newtcbox{\bluebox}{on line,
  colframe=white,colback=blue!10!white,
  height=1em,valign=bottom,
  boxrule=0.5pt,arc=2pt,boxsep=0pt,left=2pt,right=2pt,top=1pt,bottom=1pt}

\newcommand{\eat}[1]{}

\newcommand{\stitle}[1]{\smallskip\noindent\textbf{#1}}

\newtheorem{thm}{Theorem}
\newtheorem{prop}{Proposition}
\newtheorem{ex}{Example}

\newtheorem{example}[ex]{Example}
\newtheorem{definition}[thm]{Definition}

\newlength{\listingindent}                
\usepackage[LGRgreek]{mathastext}

\setlist{leftmargin=*}

\graphicspath{ {images/} }
\setcopyright{none} 

\begin{document}

\newcommand{\total}[0]{\emph{TOTAL}\xspace}
\newcommand{\cnt}[0]{\emph{COUNT}\xspace}
\newcommand{\cof}[0]{\emph{COF}\xspace}
\newcommand{\cjt}[0]{\texttt{CJT}\xspace}
\newcommand{\cjts}[0]{\texttt{CJT}s\xspace}
\newcommand{\jt}[0]{\texttt{JT}\xspace}
\newcommand{\jtivm}[0]{\texttt{JT-IVM}\xspace}
\newcommand{\jts}[0]{\texttt{JT}s\xspace}
\newcommand{\analytics}[0]{\emph{Wide-table Delta Analytics}\xspace}

\newcommand{\AnnoP}[0]{${\bf A_p}$\xspace}
\newcommand{\Anno}[0]{${\bf A}$\xspace}
\newcommand{\AnnoDiff}[0]{${\bf A_\Delta}$\xspace}
\newcommand{\BagDiff}[0]{${\bf B_\Delta}$\xspace}

\newcommand{\ewu}[1]{\red{ewu: #1\xspace}}
\newcommand{\sys}[0]{\texttt{Reptile}\xspace}
\newcommand{\matlab}[0]{\texttt{Matlab}\xspace}
\newcommand{\revise}[2]{{#1\xspace}}
\newcommand{\techreport}[1]{}

\newenvironment{myitemize}{%
\begin{itemize}[leftmargin=1em, itemsep=.1em, parsep=.1em, topsep=.1em,
    partopsep=.1em]}
{\end{itemize}}

\author{Zezhou Huang}
\email{zh2408@columbia.edu}
\affiliation{
  \institution{Columbia University}
}
\author{Eugene Wu}
\email{ewu@cs.columbia.edu}
\affiliation{
  \institution{DSI, Columbia University}
}

\begin{abstract}
Data analytics over normalized databases typically requires computing and materializing expensive joins \revise{(wide-tables)}..  Factorized query execution models execution as message passing between relations in the join graph and pushes aggregations through joins to reduce intermediate result sizes.  Although this accelerates query execution, it only optimizes a single wide-table query.  ~\revise{In contrast, wide-table analytics is usually interactive and users want to apply delta to the initial query structure. For instance, users want to slice, dice and drill-down dimensions, update part of the tables and join with new tables for enrichment. Such \analytics offers novel work-sharing opportunities.}.

This work shows that carefully materializing messages during query execution can accelerate \analytics by ${>}10^5\times$ as compared to factorized execution, and only incurs a constant factor overhead. The key challenge is that messages are sensitive to the message passing ordering. To address this challenge, we borrow the concept of calibration in probabilistic graphical models to materialize sufficient messages to support any ordering.  We manifest these ideas in the novel {\it Calibrated Junction Hypertree} (CJT) data structure, which is fast to build, aggressively re-uses messages to accelerate future queries, and is incrementally maintainable under updates.  We further show how \cjts benefit applications such as OLAP, query explanation, streaming data, and data augmentation for ML.  Our experiments evaluate three versions of the CJT that run in a single-threaded custom engine, on cloud DBs, and in Pandas, and show $30{\times}-10^5{\times}$ improvements  over state-of-the-art factorized execution algorithms on the above applications.

\end{abstract}

\title{Calibration: A Simple Trick for Wide-table Delta Analytics}
\maketitle
\pagestyle{plain}

\section{Introduction}

Schema normalization is a foundational concept in databases, and is used to minimize redundancy, potential data inconsistencies, and storage costs.  It is widely used in practice and taught in nearly every database course.  Unfortunately, normalized schemas present a number of usability challenges in modern data analytics.
First, analyses often access data from disparate tables that necessitate joining across the normalized schema (called {\it join graph}).  These massive joins are difficult to optimize~\cite{leis2018query,neumann2018adaptive,albutiu2012massively}, expensive to materialize, and dominate analytics costs.  
Second, joins are notoriously confusing to students and programmers~\cite{miedema2022identifying,kearns1997teaching,gonzalez2010google,joinshard}.

As such, there is increasing advocacy for a wide-table abstraction~\cite{widetable2,widetable2,widetable3}, where users directly perform analytics over a fully denormalized schema. 
Unfortunately, materializing a denormalized schema incurs exponential space overhead $O(n\times f^r)$, where $f$ is the fanout along edges in a join graph with $r$ relations each of size $n$.

To address this challenge, factorized query execution~\cite{schleich2019layered,aberger2017emptyheaded} accelerates queries over a large join graph using early marginalization. In the spirit of projection pushdown, early marginalization pushes down aggregation through the joins to reduce intermediate result sizes. Abo et al.~\cite{abo2016faq} established the equivalence between early marginalization and message passing in Probabilistic  Graphical Model~\cite{koller2009probabilistic} (PGM). Factorized query execution can then be modeled as passing messages between relations in the join graph. The messages are of size $O(n)$, so the space overhead (for acyclic join) is only linear: $O(r n)$.

\begin{figure}
  \centering
     \begin{subfigure}[b]{0.27\textwidth}
         \centering
         \includegraphics[width=\textwidth]{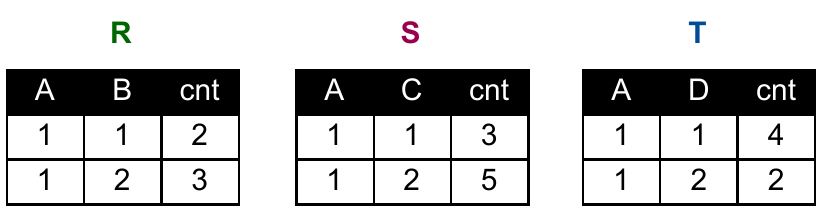}
         \caption{Relations.}
         \label{fig:relations}
     \end{subfigure}
     \hfill
    \begin{subfigure}[b]{0.2\textwidth}
         \centering
         \includegraphics[width=\textwidth]{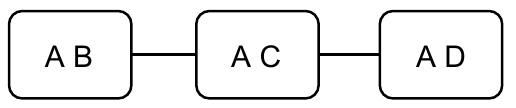}
         \caption{Join graph.}
         \label{fig:junctionHypertree}
     \end{subfigure}
     \hfill
     \begin{subfigure}[c]{0.4\textwidth}
         \centering
         \includegraphics[width=\textwidth]{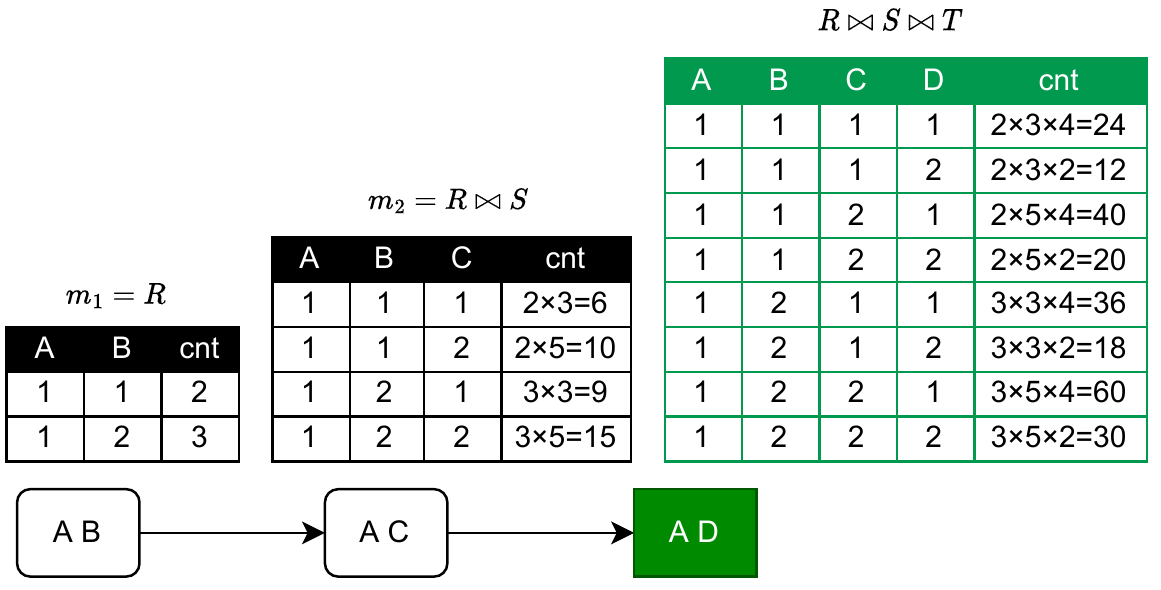}
         \caption{Naive query execution. The full join result is in green.}
         \label{fig:join}
     \end{subfigure}
     \hfill
     
     \begin{subfigure}[b]{0.45\textwidth}
         \centering
         \includegraphics[width=0.8\textwidth]{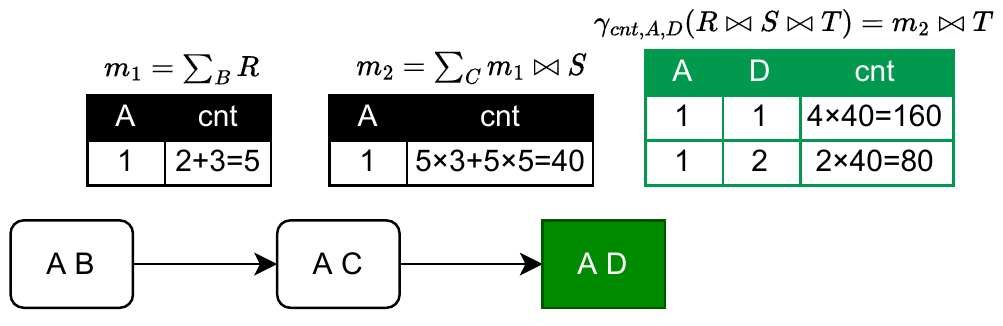}
         \caption{Upward message passing. The absorption result is in green.}
         \label{fig:messagepass}
     \end{subfigure}
     \vspace*{-3mm}
     \hfill
  \caption{Example database with three relations, its join graph (also \jt), naive query execution for total count, and factorized query execution by upward message passing.}
  \label{fig:buildDtree}
\end{figure}

\begin{example}
  \Cref{fig:buildDtree}(a,b) list example relations (duplicates are tracked with a \texttt{cnt} ``annotation'') and the join graph, respectively.  \Cref{fig:join} naively computes the total count over the full join result (wide-table) using message passing, where each message is the intermediate result so far.  For instance, \texttt{AB} sends itself to \texttt{AC}, which sends the join result to \texttt{AD}, which computes the full join before summing the counts.  This clearly requires exponential space. In contrast, factorized query execution distributes the summation through joins, so that each node first sums out (marginalizes) attributes irrelevant downstream, and then emits a smaller message.    In \Cref{fig:messagepass}, \texttt{AB} marginalizes out \texttt{B} and \texttt{AC} marginalizes out \texttt{C}, so that the final result has only 2 tuples.  

\end{example}

This concept has been extended to handle arbitrary join graph structures~\cite{olteanu2015size}; identify good message passing orders~\cite{schleich2019layered}; and support complex aggregations over semi-ring structures such as factorized learning~\cite{schleich2016learning,curtin2020rk}, where the aggregation function trains a ML model. These properties make factorized execution promising for wide-table analytics.

However, common wide-table analytics starts with an initial ({\it pivot}) query, and then applies deltas to the query structure \revise{(delta queries)}. ---these may modify selection or grouping clauses, update or remove tables, or join new tables. Further, these  use cases demand interactive response times~\cite{fisher2012exploratory,staniak2019landscape}.  For instance, users may incrementally slice, dice and drill down along dimensions~\cite{gray1997data}; apply predicate-based deletion interventions on the input tables to understand their effects~\cite{roy2015explaining,wu2013scorpion}; or join with new tables as part of ML augmentation~\cite{chepurko2020arda} or data enrichment~\cite{dong2022table}.
Although factorized execution reduces individual query latencies, it does not leverage materialization during query processing to exploit work-sharing across them (either from the pivot query, or between pre-computed data structures).  \textbf{ Our experiments show that \analytics can be ${>}10^5\times$ slower than need be}.

Our core question is: \textbf{what intermediates should be materialized for \analytics? }
First, which messages can be shared between the pivot query and delta queries? 
\revise{Delta queries only differ from the pivot query in a subset of selection, projection, join or aggregation clauses, and most messages from the pivot query could be re-used.}.
To this end, we introduce an efficient method to check the equivalence of intermediate messages.

Second, given the pivot query, which of its messages are sufficient to support re-use for \analytics?
The key challenge is that simply caching the messages emitted when executing the pivot query is insufficient because message re-usability is sensitive to the message-passing order.  This would force future delta queries to either use the same ordering or forgo message re-use.  
To address this, we observe that query execution passes messages across all edges in the join graph along a single direction, and show that sending and materializing messages in reverse for the pivot query are sufficient to support arbitrary orderings in future delta queries. 
We relate this to {\it calibration} in probabilistic graphical models~\cite{shafer1990probability}, which  similarly shares computation between interactive queries over posterior distributions. 

Third, we bring these ideas together in the \textbf{Calibrated Junction HyperTree} (\cjt).  Given a pivot query, the novel data structure manages message materialization and re-use.  Building the data structure only takes up to twice the time as executing the pivot query, but supports re-use for arbitrary message passing orders.

Finally, we apply \cjt to three important classes of wide-table applications.
\cjt accelerates \textbf{Data Cube} construction by re-using messages from low-dimensional cuboids to answer higher dimensional OLAP queries, and avoiding the exponential cost of constructing high dimensional cuboids directly.  
\cjt enables interactive \textbf{Data Augmentation for ML} by reducing the time to add a new relation (features) to the join graph and update an ML model by $>10^2\times$ compared to prior approaches.
\cjt can directly use Factorized IVM~\cite{ahmad2012dbtoaster,nikolic2018incremental} to accelerate \textbf{Data Explanation and Streaming} applications, and we further reduce IVM maintenance overheads by ${>}92\times$ by lazily maintaining messages.


\smallskip\noindent
To summarize, our contributions are as follows:

\begin{myitemize}
  \item Conceptually, we expand the connection between factorized queries and PGM by drawing on the idea of calibration.  
  \item Practically, we design the novel \cjt data structure, which uses calibration to enable work-sharing for \analytics.
	The cost of materializing the data structure is within a constant factor of the pivot query execution, but accelerates future queries by multiple orders of magnitude.
  \item We apply \cjt to data cube, data augmentation for ML, streaming and explanation applications, and describe additional application-specific optimizations.
  \item To illustrate the algorithmic benefits of our ideas, we implement and evaluate three versions of \cjt: a custom single threaded query engine and middleware compilers to SQL and Pandas data frame operations.  Our custom engine out-performs the state-of-the-art LMFAO factorized engine by ${\sim}30\times$ on OLAP queries; compared to factorized execution algorithms, our SQL experiments on AWS Redshift reduce execution by up to $10^3\times$ on TPC-H queries, while Pandas experiments accelerate data augmentation for ML by ${>}100\times$.
\end{myitemize}

\section{Background}
\label{s:background}

This section provides a brief overview of annotated relations, early marginalization and variable elimination to accelerate join-aggregation queries, and the junction hypertree join representation.  
Our goal in this paper is to keep the content accessible.  To this end, we avoid technical concepts (e.g., hypergraphs) that are needed for deriving bounds but not needed for developing intuition, and limit the discussion to \cnt queries.  However, our work generalizes to any commutative semi-ring aggregation query~\cite{green2007provenance}, and the full technical details can be found in the technical report~\cite{techreport}.

\stitle{Data Model.}  
Let uppercase symbol $A$ be an attribute, $dom(A)$ is its domain, and lowercase symbol $a\in dom(A)$ be a valid attribute value. By default, we assume categorical attributes. Numerical attributes are usually part of the semi-ring annotation discussed below. However, we can easily support numerical attributes by introducing a domain with infinite size. Given relation R, its schema $S_R$ is a set of attributes, and its domain $dom(R) = \times_{A \in S} dom(A)$ is the Cartesian product of its attribute domains. An attribute is incident of R if $A \in S_R$. Given tuple t, let $t[A]$ be its value of attribute A.

\stitle{Annotated Relations.} 
Since relational algebra (first-order logic) does not support aggregation, it has been extended with the use of 
commutative \xspace structures to support aggregation.  The main idea is that tuples are annotated with values 
from a semi-ring, and when relational operators (e.g., join, project, group-by) concatenate or combine
tuples, they also multiply or add the tuple annotations, such that the final annotations correspond to the
desired aggregation results.

A commutative semi-ring $(D, +, \times, 0, 1)$ defines a set $D$, binary operators + and $\times$ closed over $D$ where both are commutative, and the zero $0$ and unit $1$ elements.  For simplicity, the text will be based on \cnt queries and the natural numbers semi-ring $(\mathbb{N},+,\times,0,1)$, which operates as in grade school math.  However, our work extends to arbitrary commutative semi-ring structures that support aggregation queries containing common statistical functions (mean, min, max, std), as well as machine learning models (e.g., linear regression, regression trees, etc).  Our applications and experiments illustrate these use cases, and the technical report presents a full treatment~\cite{techreport}.
Each relation $R$ annotates each of its tuples $t\in dom(R)$ with a natural number, and $R(t)$ refers to this annotation for tuple $t$~\cite{green2007provenance,joglekar2015aggregations,nikolic2018incremental}.
We will use the terms {\it relation} and {\it annotated relation} interchangeably.

\stitle{Semi-ring Aggregation Query.}
Aggregation queries are defined over annotated relations, and the relational operators are extended to add or multiple tuple annotations together, so that the output tuples' annotations are the desired aggregated values\footnote{Note that this means different aggregation functions are defined over different semi-ring structures, and our examples will focus on \cnt queries.}.

Consider the query $\varphi_{\textbf{S'}} = \gamma_{\textbf{S} - \{A\},\cnt}(R_1 \Join R_2 ... \Join R_n)$
that joins $n$ relations, groups by all attributes except $A$: $\textbf{S'} = \textbf{S} - \{A\}$, and computes the \cnt.
The operators that combine multiple tuples are join and groupby (projection under set semantics corresponds to groupby), and they compute the output tuple annotations as follows:
\begin{align}
  (R\Join T)(t) =& \HS R(\pi_{S_R} (t)) \times T(\pi_{S_T} (t)) \\
  (\sum_A R)(t) = & \sum \{R(t_1) | \HS t_ 1 \in D_S , t = \pi_{S_R \backslash  \{A\}} (t_1 )\} 
\end{align}

\noindent The first statement states that given a join output tuple $t$, its annotation is defined by multiplying the annotations of the pair of input tuples, where $S_R$ and $S_T$ are $R$ and $T$'s schemas.
The second defines the count for output tuple $t$, and $\sum_{A}R$ denotes that we {\it marginalize} over $A$ and remove it from the output schema.  This corresponds to summing the annotations for all input tuples that are in the same group as $t$.

To summarize, join and groupby correspond to $\times$ and $+$, respectively, and the query can be rewritten as $\varphi_{\textbf{S'}} = \sum_{A \in  \textbf{S'}}(R_1 \Join R_2 ... \Join R_n)$.  This lets us distribute summations across multiplications as in simple algebra, as we discuss next.

\stitle{Early Marginalization.}
In simple algebra (as well as semi-rings), multiply distributes over addition, and can allow us to push marginalization 
through joins, in the spirit of projection push down~\cite{gupta1995aggregate}.  

Consider \Cref{fig:buildDtree}, which computes $\gamma_{A;\cnt}\left(  R\Join S\Join T\right)$.
We can rewrite it as marginalizing $B$, $C$, and $D$ from the full join result 
$$\sum_B \sum_C \sum_D R[A,B] \Join S[A,C] \Join T[A,D].$$
Although the naive cost is $O(n^3)$ where $n$ is the cardinality of relations, we can push down the marginalizations to derive 
the following, where the largest intermediate result, and thus the join cost, is $O(n)$:
$$\sum_D (\blue{\sum_C (}\red{(\sum_B R[A,B])} \blue{\Join S[A,C])} \Join T[A,D])$$

\stitle{Join Ordering and Variable Elimination.}
Variable elimination is a class of query execution plans that combines early marginalization with join ordering.
Early marginalization is applied to a given join order, thus we may also reorder the joins to cluster relations that involve a given attribute, so that it can be safely marginalized.
Consider the query $\sum_{A} R[A,B] \Join S[B,D] \Join T[A,C]$.
We can reorder the joins so that $A$ can be marginalized out earlier: 
$$S[B,D] \Join \sum_{A} (R[A,B] \Join T[A,C]).$$

The above procedure, where for each marginalized attribute $A$, we first cluster and join relations incident to $A$, and then marginalize $A$, is called variable elimination~\cite{cozman2000generalizing} and is widely used for inference in Probabilistic Graphic Models~\cite{koller2009probabilistic}.  The order in which attribute(s) are marginalized out (by clustering and joining the incident relations) is called the {\it variable elimination order}. Note that a given order is simply an execution plan. The complexity of variable elimination is dominated by the intermediate join result size of the clustered relations (using worst-case optimal join~\cite{ngo2018worst}). It is well known that finding the optimal order (with the minimum intermediate join size) is NP-hard~\cite{fischl2018general}. Prior work~\cite{abo2016faq} has shown that the intermediate result size of optimal order is bound by the fractional hypertree width of the join graph, which we introduce in \Cref{complexityAnalysis}. However, common database queries are over acyclic join graph, whose optimal order could be found efficiently as discussed below.

\stitle{Junction Hypertree.} 
The Junction Hypertree\footnote{\jt is also called Hypertree Decomposition~\cite{abo2016faq,joglekar2016ajar}, Join Tree, Join Forest~\cite{idris2017dynamic,schleich2019layered} in databases and Clique Hypertree in probalistic graphical models~\cite{koller2009probabilistic}.}
(\jt) is a representation of a join query that is amenable to complexity analysis~\cite{abo2016faq,joglekar2016ajar} and semi-ring aggregation query optimization~\cite{aberger2017emptyheaded}. In the next section, we will show how \jt  can be materialized and maintained to provide work sharing and optimization opportunities for join-aggregation queries, and how to extend it to balance storage costs and query benefits.    For now, we simply define the structure.

Given a join graph $R_1\Join \ldots \Join R_n$ using natural joins for simplicity, a Junction Hypertree is a pair 
$(E, V)$, where each vertex $v\in V$ is a subset of attributes in the join graph, and the undirected edges form a tree that spans the vertices.  
The join graph may be explicitly defined by a query, or induced by the foreign key relationships in a database schema.
Following prior work~\cite{abo2016faq}, a \jt vertex is also called a {\it bag}.
A \jt must satisfy three properties:
\begin{itemize}
  \item \stitle{Vertex Coverage:} the union of all bags in the tree must be equal to the set of attributes in the join graph.  
  \item \stitle{Edge Coverage:} for every relation $R$ in the join graph, there exists at least one bag that is a superset of $R$'s attributes.
  \item \stitle{Running intersection:} for any attribute in the join graph, the bags containing the attribute must form a connected subtree.  In other words, if two bags both contain attribute $A$, all bags along the path between them should also contain $A$.  
\end{itemize}

The last property is important because \jts are related to variable elimination and are used for query execution.  Given an elimination ordering, let each join cluster be a bag in the \jt, and adjacent clusters be connected by an edge.  In this context, executing the variable elimination order corresponds to traversing the tree (path); when execution moves beyond an attribute's connected subtree, then it can be safely marginalized out.  
Note that since the \jt is undirected, it can induce many variable elimination orders (execution plans) from a given \jt, all with the same runtime complexity.

Finally, there are many valid \jt for a given join graph, and the complexity (query execution cost) of a \jt is dominated by the largest bag (the join size of the relations covered by the bag).  Although finding the optimal \jt for an arbitrary join graph is NP-hard~\cite{fischl2018general}, we can trivially create the optimal \jt for an acyclic join graph by creating one bag for each relation (e.g., the \jt is simply the join graph) and the size of each bag is bounded by its corresponding relation size. We refer readers to FAQ~\cite{abo2016faq} for a complete description. 

\begin{figure}
  \centering
  \begin{subfigure}[t]{0.23\textwidth}
         \centering
         \includegraphics[width=\textwidth]{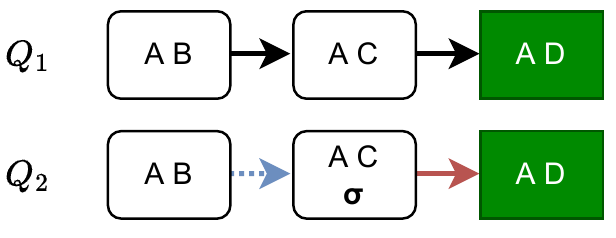}
          \caption{Message passing to root AD.}
          \label{fig:sharebetweenqueries}
     \end{subfigure}
     \hfill
    \begin{subfigure}[t]{0.23\textwidth}
         \centering
         \includegraphics[width=\textwidth]{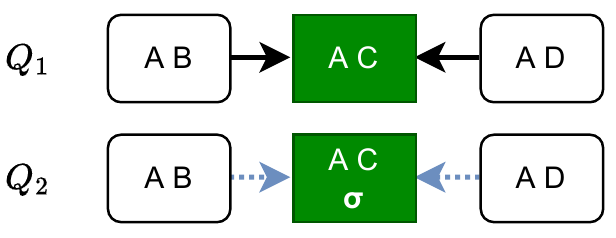}
         \caption{Moving the root increases message reuse.}
          \label{fig:sharebetweenqueries2}
     \end{subfigure}
     \hfill
     \vspace*{-3mm}
      \caption{Work sharing opportunities between queries $Q_1$ (total count query) and $Q_2$ with additional predicate applied to S(A,C). Dotted \blue{blue edges} are reusable messages and solid \red{red edges} are non-reusable edges. }
\end{figure}

\stitle{Message Passing for Query Execution.}
Message Passing was first introduced by Judea Pearl in 1982~\cite{pearl1982reverend} (known as belief propagation) in order to efficiently perform inference (compute marginal probability) over probabilistic graphical models. 
In database terms,  each probability table corresponds to a relation, the probabilistic graphical model corresponds to the full join graph in a database (as expressed by a \jt), the joint probability over the  model  corresponds to the full join result, and marginal probabilities correspond to grouping over different sets of attributes.
To further support semi-ring aggregation, Abo et al.~\cite{abo2016faq} established the equivalence between variable elimination,  factorized execution of a single query, and (upward) message passing.
Below, we illustrate how message passing over a \jt is used for query execution, and the next section leverages the ability to reuse messages across queries.

The procedure first determines a traversal order over the \jt---since the \jt is undirected, we can arbitrarily choose any bag as the root and create directed edges that point towards the root---and then traverses from leaves to root.  We first compute the initial contents of each bag by joining the necessary tables based on the bag's attributes.  When we traverse an outgoing edge from a bag $l$ to its parent $p$, we marginalize out all attributes that are not in their intersection---the result is the {\it Message} between $l$ and $p$.  The parent bag then joins the message with its contents.  Each bag waits until it has received messages from all incoming edges before it emits along its outgoing edges, and once the root has received all incoming messages, its updated contents correspond to the query result.

\begin{example}[Message Passing]
Consider the relations in \Cref{fig:relations}, and the \jt in \Cref{fig:junctionHypertree} where each bag is a base relation.
  We wish to execute $\sum_{ABCD} R(A,B) \Join S(A,C) \Join T(A,D)$ by 
  traversing along the path $AB\to AC\to AD$ (\Cref{fig:messagepass}).
  We first marginalize out $B$ from $AB$, so the message to $AC$ is a single row with count $5$.   
  The bag $AC$ joins the row with its contents, and thus multiplies each of its counts by $5$.
  It then marginalizes out $C$, so its message to $AD$ is a single row with count $(3+5)\times 5$.
  Finally, bag $AD$ absorbs the message (\Cref{fig:messagepass}) and marginalizes out $A$ and $D$ to compute the final result.
\end{example}

\stitle{Scope.} Following prior factorized query execution work~\cite{schleich2019layered,nikolic2018incremental,olteanu2015size}, we use the acyclic join graph as the \jt, or assume a good \jt has already been determined from standard hypertree decomposition ~\cite{abo2016faq,joglekar2016ajar} for cyclic join graphs.  Although \cjt supports any commutative semi-ring, theta joins with arbitrary conditions, outer joins, and any factorized machine learning model\footnote{Including ridge regression, classification tree, regression tree~\cite{schleich2019layered}, k-means (RK-means), support vector machine~\cite{khamis2020functional} and factorization machine~\cite{schleich2019layered}.}, we try keep the ideas accessible to a general database audience and base our examples on natural numbers (a semi-ring), \cnt queries, with natural joins, and the linear regression model.

\section{Calibrated Junction Hypertree}
\label{cjt_detail}
While message passing over \jt exploits early marginalization to accelerate query execution, it has traditionally been limited to use in complexity analysis and single-query execution. 
This section introduces the Calibrated Junction Tree (\cjt) to enable work-sharing for efficient \analytics.  
The idea is to materialize messages over the \jt for the  {\it wide-table pivot query}, and reuse a subset of its messages for future {\it delta queries}.  
This section will focus on the basis for the \cjt data structure and how it is used to execute {\it delta queries}. The next section will describe how to customize \cjt for a range of useful wide-table applications.  

Our novelty is 1) to use \jts as a concrete data structure to support message reuse across queries, and 2) to borrow {\it Calibration}~\cite{shafer1990probability} from probabilistic graphical models to ensure that sufficient messages are materialized to efficiently support arbitrary SPJA queries over the full join graph.   Although \cjt is widely used across engineering~\cite{zhu2015junction,ramirez2009fault}, ML~\cite{braun2016lifted,deng2014large}, and medicine~\cite{pineda2015novel,lauritzen2003graphical}, we are the first to introduce \cjt in the context of query execution and generalize it to semi-ring aggregation queries.

\subsection{Motivating Example}
\label{sec:mot}

We first illustrate work sharing examples between a pivot query 
$Q_1 = \sum_{ABCD} AB \Join AC \Join AD$ (whose messages are materialized) and 
a delta query $Q_2 = \sum_{ABCD} AB \Join \sigma_{c=1}(AC) \Join AD$.
$Q_1 $ computes the total count, and $Q_2$ applies an additional predicate \texttt{C=1}.  

\begin{example}
  The \jts in \Cref{fig:sharebetweenqueries} both assign $AD$ as the root and traverse along the path $AB\to AC\to AD$. 
  Although the message $AB\to AC$ will be identical (\blue{blue} edges), the additional filter over $AC$ means that its outgoing message (and all subsequent messages) will differ from $Q_1$'s and cannot be reused (\red{red} edges).
In contrast, \Cref{fig:sharebetweenqueries2} uses $AC$ as the root, so both messages can be reused and the 
  $AC$ bag simply applies the filter after joining its incoming messages.
\end{example}

This example shows that message reuse depends on how the root bag is chosen in the pivot query,  
and for different delta queries, we may wish to choose different roots.  
Since we may not know the exact join, grouping, and filter criteria of future delta queries, it is not effective to simply materialize
messages for a single root.
The \cjt data structure addresses these limitations, and the following text describes the \cjt data structure, message reuse, and query execution given a \cjt.

\subsection{Junction Hypertree as Data Structure}

A naive approach to re-use messages is to execute an aggregation query over a \jt, and store the messages; when a future query traverses an edge in the \jt, it simply reuses the corresponding message.  Unfortunately, this is 1) inaccurate, because messages generated along an edge are not symmetric, so that message contents depend on the specific traversal order during message passing, 2) insufficient, because it cannot directly express all filter-group-by queries over the \jt, and 3) leaves performance on the table.    
To do so, we extend the \jt data structure as follows:

\stitle{Directed Edges.} To support arbitrary traversal orders, we replace each undirected edge with two directed edges, and use $\mathcal{Y}(i\to j)$ to refer to the cached message for the directed edge $i\to j$.

\begin{figure}
  \centering
      \includegraphics [scale=0.7] {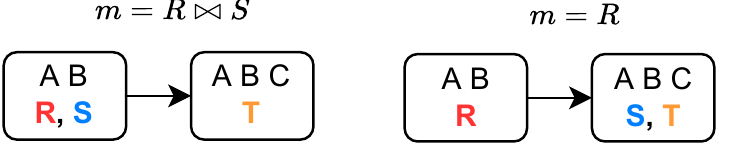}
      \vspace*{-3mm}
      \caption{ The same \jt over relations \textcolor{red}{R(A,B)}, \textcolor{blue}{S(A)}, \textcolor{orange}{T(B,C)} can have different relation mappings ($\mathcal{X}$) and each mapping results in different messages ($m$). For each bag, its attributes are at the top and mapped relations are at the bottom.}
  \label{fig:diffmsg}
\end{figure}

\stitle{Relation Mapping.} $\mathcal{X}(R)$ maps each base relation $R$ to exactly one bag containing $R$'s schema.  Although different mappings can lead to different messages (\Cref{fig:diffmsg}), acyclic join graphs have a good default mapping where each relation maps to a single bag\footnote{Heuristics for the general case have been studied by the greedy variable elimination in Probabilistic Graphical Model~\cite{koller2009probabilistic}.}.   Relations mapped to the same bag are joined during message passing.

\stitle{Empty Bags.}  To avoid large paths during message passing it can help to add custom {\it Empty} bags to create ``short cuts''.  {\it Empty} bags are not mapped from any relations and simply a mechanism to materialize custom views for work sharing.  They join incoming messages, marginalize using standard rules, and materialize the outgoing messages.   Empty bags are a novel addition in this work:  previous works~\cite{abo2016faq,aberger2017emptyheaded,xirogiannopoulos2019memory} focus on non-redundant \jt without empty bags in the context of single query optimization.

\begin{figure}
  \centering
  \begin{subfigure}[t]{0.23\textwidth}
         \centering
         \includegraphics[width=0.8\textwidth]{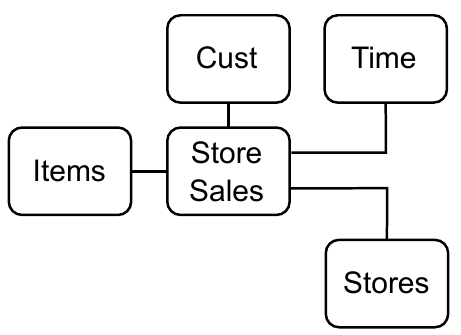}
          \caption{TPC-DS join graph (also \jt)}
          \label{fig:tpc_ds_schema}
     \end{subfigure}
     \hfill
    \begin{subfigure}[t]{0.23\textwidth}
         \centering
         \includegraphics[width=0.8\textwidth]{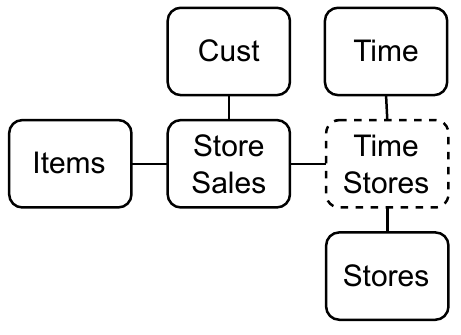}
         \caption{Add empty bag (Time, Stores).}
          \label{fig:tpc_ds_empty}
     \end{subfigure}
     \hfill
     \vspace*{-3mm}
      \caption{The join graph and \jt of TPC-DS (simplified). Adding an empty bag can accelerate queries group-by \texttt{Time} and \texttt{Stores}.}
\end{figure}

\begin{example}[Empty Bag]
  Consider the simplified TPC-DS \jt in \Cref{fig:tpc_ds_schema}. Store Sales is a large fact table (2.68M rows at SF=1), while the rest are much smaller.  To accelerate a query that aggregates \texttt{sales} grouped by \texttt{(Store,Time)}, we can create the empty bag \texttt{Time Stores} between \texttt{Store\_Sales}, \texttt{Time} and \texttt{Stores} (\Cref{fig:tpc_ds_empty}). The message from \texttt{Store\_Sales} to the empty bag is sufficient for the query and is $17.3\times$ smaller (154K rows) than the fact table.
\end{example}

Note that leaf empty bag may result in an empty output message; we avoid this special case by mapping the identity relation $\mathcal{I}$\footnote{The schema is the same as the bag and all tuples in its domain are annotated with 1 element in the semiring.} to it, such that $R\Join \mathcal{I} = R$ for any relation $R$. Essentially, the empty bag is ``pass-through'' and doesn't change the join results nor the query result. When the bag is a leaf node, its message is simply $\mathcal{I}$.  We do not materialize the identity relation, as it's evident from the \jt.

\begin{figure}
  \centering
     
    \begin{subfigure}[c]{0.2\textwidth}
         \centering
         \includegraphics[width=\textwidth]{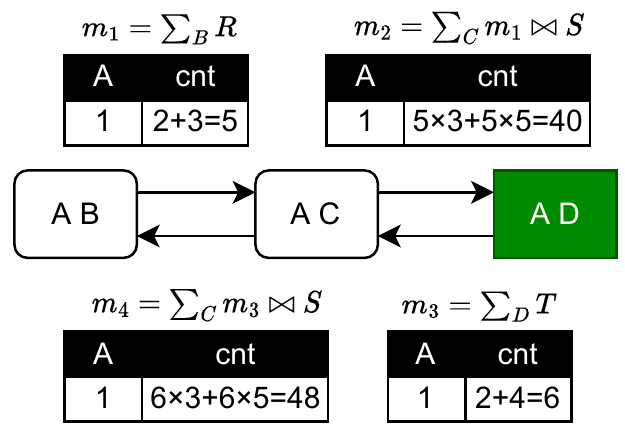}
         \caption{Upward and downward message passing. Green rectangle is the root.}
         \label{fig:downward}
     \end{subfigure}
     \hfill
     \begin{subfigure}[c]{0.24\textwidth}
         \centering
         \includegraphics[width=\textwidth]{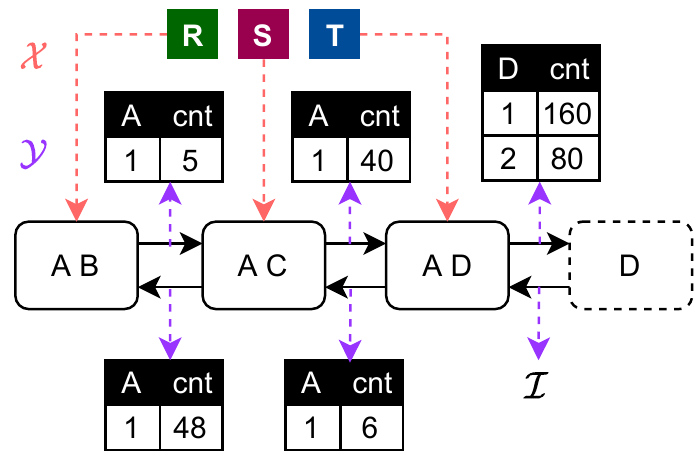}
         \caption{Calibrated Junction Hypertree with empty bag (dotted). $\mathcal{I}$ is the identity relation.}
         \label{fig:datastructure}
     \end{subfigure}
     \hfill
     \vspace*{-3mm}
  \caption{Message Passing and Calibration}
\end{figure}

\begin{example}[\jt Data Structure]
 \Cref{fig:datastructure} illustrates the \jt data structure for the example in \Cref{fig:buildDtree}. Each relation maps to exactly one bag (orange dotted arrows), and each directed edge between bags (black arrows) stores its corresponding message (purple dashed arrows).  Bag D (dotted rectangle) is an empty bag and materializes the view of "count group by D".  $\mathcal{I}$ is the identity relation.
\end{example}

\begin{table*}
\begin{center}
\setlength{\tabcolsep}{0.4em} 
\begin{tabular}{  m{5em}  m{30em} m{12em} l} 
  \textbf{Annotation} & \textbf{Effect} & \textbf{Applicability} & \textbf{Section}   \\ 
  $\gamma_A$ & Prevent A from being marginalized out for all downstream messages. & Any bag containing A. & \Cref{sec:msgpassing}  \\ 
  $\sum_A$ & Marginalize out $A$.  ``Cancels'' $\gamma_A$ for downstream messages. & Any bag containing A. & \Cref{sss:cjtqexec} \\ 
$\overline{R}$ & Exclude relation R from the bag during message passing. & The bag $\mathcal{X}(R)$.  & \Cref{sec:msgpassing}\\ 
\revise{$R^*_{ver.}$}. & Update relation R in the bag to the specified version during message passing. & The bag $\mathcal{X}(R)$.  & \Cref{sec:msgpassing}\\ 
$\sigma_{id}$ & Apply selection specified by id to relations during message passing.  & Any bag $\sigma_{id}$ is applied to.& \Cref{sec:msgpassing}  \\ 
\end{tabular}

\end{center}
\caption{Table of annotations, their effects and applicability.}
\vspace*{-7mm}
\label{table:annotations}
\end{table*}

\subsection{Message Passing Over Annotated Bags}
\label{sec:msgpassing}

We now describe support for general SPJA queries over \jt.  Although each query \jt has the same structure, we annotate the bags based on the query's SPJA operations. We then modify message passing rules to accommodate the bag annotations.    These annotations will come in handy when determining work sharing opportunities for a new delta query given a pivot query.  

Given the database  $\textbf{R}= \{R_1,R_2,...,R_n\}$ and \jt $ = ((E,V), \mathcal{X}, \mathcal{Y})$, we focus on queries of the following form, where any semi-ring aggregation is acceptable:\\

\indent\texttt{SELECT \textcolor{red}{$\mathcal{G}$}, COUNT(*)} \texttt{FROM \textcolor{blue}{$\mathcal{J}$}}\\
\indent\texttt{WHERE [JOIN COND] AND \textcolor{orange}{$\mathcal{P}$}} 
\texttt{GROUP BY \textcolor{red}{$\mathcal{G}$}}\\

\noindent where $\mathcal{G}$ is the grouping attributes, $\mathcal{J}\subseteq \mathcal{R}$ is the set of relations joined in the \texttt{FROM} clause, and $\mathcal{P}$ is the set of single-attribute predicates\footnote{Multi-attributes predicates have interesting optimization opportunities~\cite{khamis2020functional} but is not our focus. We rewrite them into group-by those attributes followed by the predicate. }. 
Query execution is based on message passing as in \Cref{s:background}, however the processing at each bag differs based on its annotations.
We propose 4 annotation types, summarized in \Cref{table:annotations}:

\begin{myitemize}

\item \texttt{\textcolor{red}{GROUP BY $\mathcal{G}$.}} For each attribute $A\in\mathcal{G}$, we annotate exactly one bag $u$ that contains this attribute with $\gamma_A$.  Messages emitted by the annotated bag and all downstream bags do not marginalize out $A$.  Since all bags containing $A$ form a connected subtree, which bag we annotate does not affect correctness,  however we will later discuss the performance implications of different choices when we use the annotated \jt to execute delta queries.

\item \texttt{\textcolor{blue}{Joined Relations $\mathcal{J}$.}} The query may not join all relations  in the join graph, \revise{or the joined relations are updated.}.
  For each relation R {\it not} included (resp. updated) in the query, we annotate the corresponding bag $u=\mathcal{X}(R)$  with $\overline{R}$ (resp. $R^*_{ver.}$). When computing messages from this bag, R will be excluded from $\mathcal{X}^{-1}(u)$\footnote{Rigorously, $\mathcal{X}$ doesn't have an inverse function. We define $\mathcal{X}^{-1}$ to be a mapping from one bag to a set of base relations such that $\mathcal{X}^{-1}(u) = \{i| \mathcal{X}(i) = u\}$.} (resp. R will be updated in $\mathcal{X}^{-1}(u)$). We allow only exclusions of relations that don't violate \jt properties. \revise{The query could further apply IVM or join with new relations as discussed in \Cref{app:olap}}..

\item \texttt{\textcolor{orange}{PREDICATES $\mathcal{P}$.}} Let predicate $\sigma\in\mathcal{P}$ be over attribute $A$.  
  If $A$ is not in any relation in \textcolor{blue}{$\mathcal{J}$}, we can skip it.  
  Otherwise, we choose a bag $u$ that contains $A$, and annotate it with $\sigma_{id}$---the effect is that the predicate filters all messages emitted by $u$.   
  The choice of bag to annotate is important---for a single query, we want to pick a bag far from the root in the spirit of selection push down, whereas to maximize message re-usability, we want to pick the bag near the root.  We discuss this trade-off in the next section.

\end{myitemize}

\subsubsection{Message Passing}
\label{sss:msgpassing}

We now modify how message passing, generation, and absorption work to take the annotations into account.

\stitle{Upward Message Passing.} 
Traditional message passing chooses a root bag and traverses edges from leaves to the root.  Since the \jt data structure uses bidirectional edges, we call this procedure upward message passing, as it materializes messages along edges that point towards the root.

\stitle{Message Generation $\mathcal{Y}(b{\to}p)$.}  
The message $\mathcal{Y}(b{\to}p)$ from bag $b$ to parent $p$ is defined as follows.
Let $M(b) = \{\mathcal{Y}(i{\to}b) | i{\to}b\in E\land i\not\eq p\}$ be the set of incoming messages (except from $p$).
We join between all  relations (updated to the specified versions) in $M(b)$ and $\mathcal{X}^{-1}(b)$, and marginalize out all attributes not in $p$.
Given annotations, we exclude relations in $\overline{R}$ from the join, apply predicates $\sigma$ (with appropriate push-down), and exclude attributes in $\gamma$. 
$$\mathcal{Y}(b\to p) = \sum_{b-(p\cap b)-\gamma} \sigma\left(\Join (M(b) \cup \mathcal{X}^{-1}(b) - \overline{R})\right) $$

\noindent $b$'s message to $p$ is ready iff all its messages from child bags are received.
During message passing, if $b$ contains group-by annotation $\gamma$,  we temporarily annotate all its parent bags also with $\gamma$.

\stitle{Absorption.} 
Absorption is when the root bag $r$ consumes {\it all} incoming messages.
It is identical to the join and filter during message generation:
$Absorption(r) = \sigma\left(\Join (M(r) \cup \mathcal{X}^{-1}(r) - \overline{R})\right)$ where $M(r) = \{\mathcal{Y}(i{\to}r)|i{\to}r\in E\}$.
To generate the final query results, we marginalize away all attributes not in the query's grouping conditions $\mathcal{G}$.  

\begin{figure}
  \centering

         \includegraphics[width=0.4\textwidth]{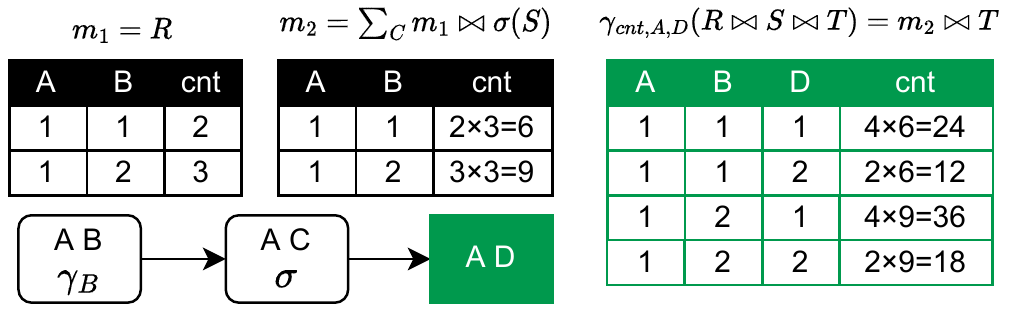}
         \vspace*{-3mm}
         \caption{Filter-group-by query with annotated \jt.}
         \label{fig:group_by_filter_query}

\end{figure}

\begin{example}
Consider database and \jt in \Cref{fig:buildDtree}. Suppose we want to query the total count filter by C = 1 and group by B. This requires us to annotate bag AB with $\gamma_B$ and bag AC with $\sigma$ (id is omitted). \Cref{fig:group_by_filter_query} shows the upward message passing over the annotated \jt to root AD, where attribute B is not marginalized out and the predicate C = 1 is applied to S. After upward message passing, bag AD performs absorption and marginalizes out AD to answer the query. 
\end{example}

\stitle{Single-query Optimization.} For a given SPJA query, we can choose different bags to annotate, and different roots for upward message passing.  We make these choices based on heuristics that minimize the worst-case query complexity.  Since relation removal annotation $\overline{R}$ and update anntation $R^*_{ver.}$ can be only placed on $\mathcal{X}(R)$, and the placement of group-by don't affect the message passing, the only factor is the choice of root bag and selection annotations.  We enumerate every possible root bag, greedily push down selections, and choose the root with the smallest complexity; the total time complexity to find the root is polynomial in the number of bags.

\subsubsection{Message Reuse Across Queries} 
\label{sss:msgreuse}
Messages reuse between queries requires that the message along edge $u\to v$ only depends on the annotated sub-tree rooted at $u$.  Thus, a new query can reuse materialized messages in the pivot query's \jt that have the same subtree (and annotations).  

\begin{prop}[Message Reusability]
\label{prop:messagereuse}
Given a \jt and annotations for two queries, consider directed edge $u\to v$ present in both queries.  Let $T_u$ be the subtree rooted at $u$. If the annotations for $T_u$ are the same  for both queries, then the message along $u\to v$ will be identical irrespective of the traversal order nor  choice of the root.
\end{prop}

This proposition is well established in probabilistic graphical models~\cite{shafer1990probability}, and follows for message passing over \jt.    The proof sketch is as follows: leaf nodes send messages that only depend on its outgoing edges, base relations and annotations, while a given bag's outgoing message only depends on its mapped relations in $\mathcal{X}$, annotations and its incoming messages. None of these messages depend on the traversal order nor the root.

\begin{figure}
  \centering

         \includegraphics[width=0.3\textwidth]{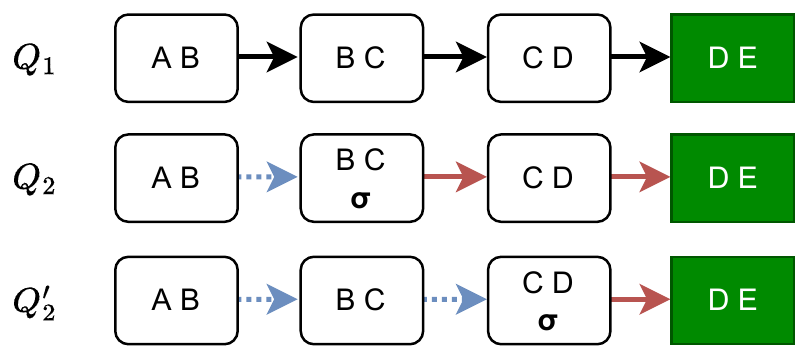}
         \vspace*{-3mm}
         \caption{Message size vs reuse trade-off. Given total count query $Q_1$, $Q_2$ adds a predicate to C. Pushing down selection as in $Q_2$ may reduce message size but hinders reuse as compared to $Q'_2$. Dotted \blue{blue edges} are reusable messages and solid \red{red edges} are non-reusable edges.}
         \label{fig:selection_pushdown}

\end{figure}

\Cref{prop:messagereuse} implies that an annotation can ``block'' reuse along all of its downstream messages. For group-by annotation, we greedily push down it to the leaf of the connected subtree closest to the root to maximize reusability.
However, pushing selections down trades-offs potentially smaller message sizes for limited reusability:

\begin{example}
  Suppose we have materialized messages for $Q_1$ in \Cref{fig:selection_pushdown}, and want to execute $Q_2$, which has an additional predicate over C.  If we annotated bag \texttt{BC} with $\sigma$, this may reduce the message size but we cannot reuse the message in $\texttt{BC} \to \texttt{CD}$. If we annotate \texttt{CD} ($Q'_2$), we can reuse the message but risk larger message sizes. 
\end{example}

In practice, we prioritize reuse by pulling annotations close to the root---reuse helps avoid scan, join, and aggregation costs, whereas larger message sizes simply increase scan sizes.

\subsection{Calibration}
\label{sec:calibration}
We saw above that message reuse depends on choosing a good root for message passing, however upward message passing only materializes messages for a single root bag.  \textit{Calibration} materializes messages along edges in the opposite direction, and thus lets future delta queries pick arbitrary roots.

\subsubsection{Calibration} 
Given an edge $u\to v$, $u$ and $v$ are calibrated iff their marginal absorption results are the same in both directions: 
$$\sum_{u - (v \cap u)}  Absorption(u) = \sum_{v - (v \cap u)} Absorption(v)$$
The \jt is calibrated if all pairs of adjacent bags are calibrated. We call this a {\it Calibrated Junction Hypertree} (\cjt), which is achieved by Downward Message Passing.

\stitle{Downward Message Passing}. Upward message passing computes messages along half of the edges (from leaves to root). Calibration simply reverses the edges and runs upward message passing from root (now the leaf) to leaves (now all roots).  Now, all edges store materialized messages.  Our algorithm extends Shafer–Shenoy inference algorithm~\cite{shafer1990probability} in PGM to semiring aggregation, and is fully described in the technical report~\cite{techreport}.  If the semi-ring is also a semi-field, the Hugin algorithm~\cite{lepar2013comparison} can further avoid redundant multiplications. This benefits common aggregation functions such as sum, stddev, and even gram matrix computation for training linear models~\cite{schleich2016learning}.

\begin{example}
Consider the example in \Cref{fig:downward}. During upward message passing, AD is the root and has received all incoming message from leaf AB. After that, we send messages back from AD to AB. We can verify that the \jt is calibrated by checking the equality between marginal absorptions.
\end{example}

Calibration means all bags are ready for absorption.
This immediately accelerates the class of queries that adds one grouping or filtering attribute $A$ to the pivot query\footnote{These queries correspond to marginal posterior probability (group-by) and incremental update (filter) in PGM~\cite{koller2009probabilistic}.}.   We simply pick a bag containing $A$ and apply the filter/group-by to its absorption result.

\stitle{Generalization of Prior Work.}
Calibration generalizes the two-pass semi-join reduction in Yannakakis's algorithm~\cite{yannakakis1981algorithms}, which is simply calibration over a 0/1 semi-ring.  Similarly, the upward-downward passes to compute local sensitivities~\cite{tao2020computing} is calibration for \texttt{COUNT} over an acyclic join graph.

\subsubsection{Query Execution Over a \cjt}
\label{sss:cjtqexec}

How do we execute a new delta query $Q$ over the \cjt of a pivot query $Q_p$?   
Since they share the same \jt structure, they only differ in their annotations.  The main idea is that query execution is limited to the subtree where the annotated bags differ between the two queries, while we can reuse messages for all other bags in the \cjt.

\begin{figure}
  \centering

         \includegraphics[width=0.45\textwidth]{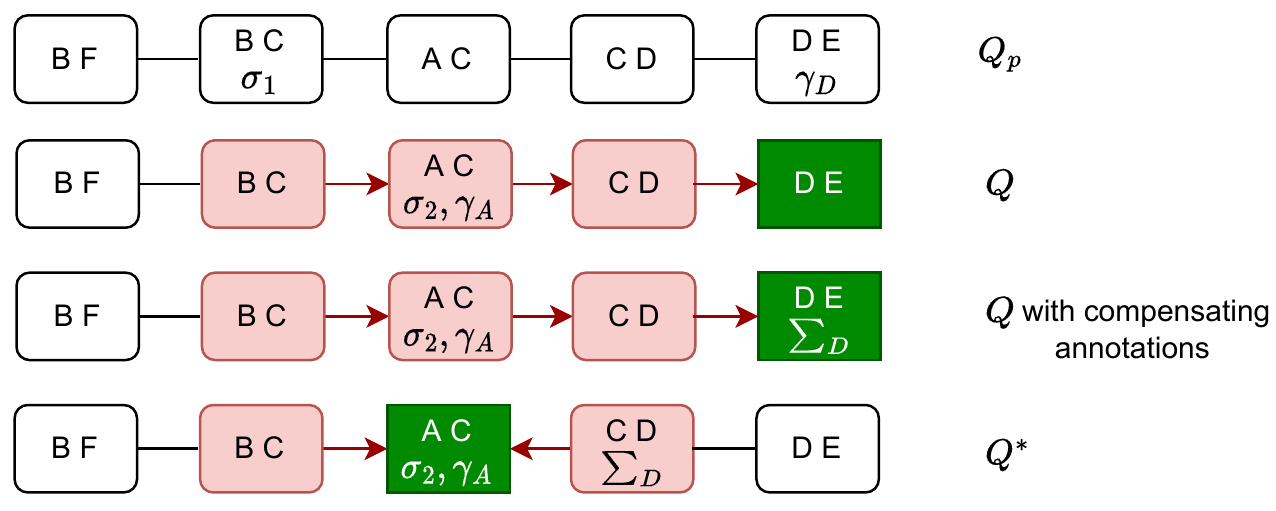}
         \vspace*{-3mm}
         \caption{Given \cjt with pivot query $Q_p$, the Steiner Tree to execute $Q$ is highlighted (green is Steiner Tree root and red is Steiner Tree non-root nodes.). $Q^*$ is the optimal query with minimum Steiner Tree size and run time complexity. }
         \label{fig:steiner_tree}

\end{figure}

Let \AnnoP and \Anno be the set of annotations for $Q_p$ and $Q$, respectively; note that the annotations in \AnnoP are bound to specific bags in the \cjt, while the annotations in \Anno are not yet bound.  Further, let \BagDiff be the subset of bags whose annotations differ between the two queries.  The steiner tree $T$ is the minimal subtree in the \cjt that connects all bags in \BagDiff.  From \Cref{prop:messagereuse}, edges that cross into $T$ have the same messages as in the \cjt and can be reused.  Thus, we only need to perform upward message passing inside of $T$.    Let us first start with an illustrative example:

\begin{example}[Steiner Tree]
  In \Cref{fig:steiner_tree}, the pivot query $Q_p$ groups by \texttt{D} and filters by $B = 1$, and so its annotations are \AnnoP $= \{\sigma_1, \gamma_D\}$. 
  Suppose query $Q$ (row 2) instead groups by A and filters by $C = 1$ so \Anno $= \{\sigma_2, \gamma_A\}$, and we place its annotations $\sigma_2$ and $\gamma_A$ on \texttt{AC}.    The two queries differ in bags \BagDiff=$\{BC, AC, DE\}$, and we have colored their steiner tree.  Naively, we can reuse the message $BF\to BC$, but otherwise re-run upward message passing along the steiner tree.
  
\end{example}

Although the example allows us to re-use one message, it is a sub-optimal execution plan because the steiner tree is not minimal, and the root is poorly chosen.  Instead, we use a greedy procedure to find the minimal steiner tree; we arbitrarily place the annotations on valid bags to create an initial steiner tree, and then greedily shrink it.  Given the minimal steiner tree, we find the optimal root following \Cref{sec:msgpassing}. 

\stitle{Initialization.}
For annotations only in \Anno, they are added to $Q$'s \jt based on the single-query optimization rules in \Cref{sec:msgpassing}.
For annotations only in \AnnoP, we need to compensate for their effects.  For $\sigma_p$ and $\overline{R}$, we remove the annotation, while for $\gamma_D$, we introduce the compensating annotation $\sum_D$, which marginalizes out $D$, and place it on the same bag.  A key property of $\sum_D$ is that we can freely place it on any bag that contains $D$.    For all of the above annotations, we add their bags to \BagDiff.  This defines the initial steiner tree ($Q$ with compensating annotations in \Cref{fig:steiner_tree}).

\begin{example}
  The third row in \Cref{fig:steiner_tree} adds the compensating annotation $\sum_D$ to \texttt{DE}.  Its execution is as follows: \texttt{BC} does not apply $B=1$, \texttt{AC}  applies $C=1$ and groups by $A$, and \texttt{DE} maginalizes out \texttt{D} and \texttt{E}.  
\end{example}

\stitle{Shrinking.}
Given the leaves of the steiner tree, we try to move their annotations towards the interior of the tree.  Recall that $\sigma$, $\gamma$, and $\sum$ can be placed on any bag containing the annotation's attribute.  We greedily choose the bag with the largest underlying relation and try to move its annotations first.  Once a leaf bag does not have any annotations that differ from the pivot \cjt, it is removed from the steiner tree.  

\begin{example}
  $Q^*$ in \Cref{fig:steiner_tree} shows the optimal execution plan over the minimal steiner tree for $Q$.   It has moved $\sum_D$ to \texttt{CD}, and made \texttt{AC}  the root.   \texttt{CD} will marginalize out \texttt{D}, and \texttt{AC} performs the filter and group-by. In this way, we also reuse the message $\texttt{DE} \to \texttt{CD}$. 
\end{example}

After the optimization procedure above, the execution plan over the \cjt will always be as efficient (or in many cases much more efficient than) executing the query without the \cjt.  Our proof sketch analyzes two scenarios.  If the optimal root without the \cjt is within the steiner tree, we can reuse messages outside the steiner tree.  If the optimal root is outside the steiner tree, we can still move the root to the closest bag within the steiner tree.    In both cases, all messages within the steiner tree are the same.  Calibration is the key mechanism that allows us to freely move the root.

\section{Applications}
\label{s:apps}

Semi-ring structures are highly expressive, and also support a wide range of \analytics.  We now describe how to choose good\footnote{In general, finding the optimal pivot query and \jt structure is NP-hard.} pivot queries and leverage \cjt for four popular classes of applications: OLAP, Explanation, Streaming and Data Augmentation for ML. We note that \cjt also applies to differential privacy~\cite{tao2020computing}, web table analysis~\cite{santos2021correlation} and what-if analysis~\cite{lakshmanan2008if}, and leave elaborations to future work.

\subsection{OLAP Data Cubes}
\label{app:olap}

OLAP data cubes~\cite{gray1997data} materialize a lattice of data cuboids parameterized by the set of attributes that future queries will filter/group by.  Traditionally, the data structure is built bottom-up in order to share computation---each cuboid is built by marginalizing out irrelevant attribute(s) from a descendant cuboid.  If the cube is over a join graph, then there is the additional cost of first materializing the (potentially very large) join result to compute the bottom cuboid.  Although prior work explored many optimizations (parallelization~\cite{wang2013scalable,dehne2002parallelizing,taniar2002parallel}, approximation~\cite{vitter1998data}, partial materialization~\cite{han1998selective,wang2002condensed}, early projection~\cite{kotsis2000elimination}), neither early marginalization nor work-sharing based on \cjts have been explored.

\cjts are a particularly good fit for building data cubes because, in practice, they are restricted to a small number of attributes in order to avoid exponentially large cuboids.    In this setting, we can build \cjts for a carefully selected set of pivot queries to accelerate cube construction by
 1) not materializing the full join graph when building the cuboids, and 2) aggressively reuse messages to answer OLAP queries not directly materialized by a cuboid.

\subsubsection{Complexity Analysis}
\label{app:cube_complexity}
Let us first analyze the complexity of using \cjts to answer OLAP queries.  This will provide the tools to trade-off between OLAP query performance and space requirements for materialization.

Let the database contain $r$ relations each with $O(n)$ rows, the domain of each attribute is $O(d)$, and the join graph contains $m$ unique attributes.  
Suppose we have calibrated each cuboid with $k$ group-by attributes (the pivot queries).  Calibration costs $O(rnd^k)$ for each pivot query where the cost per bag is $O(nd^k)$ (cross product between relation size $n$ and incoming messages). Since the output message size is also bound by $O(nd^k)$ due to marginalization, we incur this cost for each of $r$ bags in the \cjt.   Thus the total cost is $O(rn(dm)^k)$ to calibrate all $\binom{k}{m} = O(m^k)$ pivot queries.

To simplify our analysis, let us also materialize the absorption results (join result of all incoming messages and relations mapped to a given bag) for each bag during calibration (\Cref{sss:msgpassing}).  This does not change the worst-case runtime complexity, and increases the storage cost by at most the size of the base relations.

Notice that these absorption results can be directly used to answer OLAP queries with $k+1$ attributes with no cost in complexity (\Cref{sss:msgreuse}).  Thus, materializing cuboids of up to $k+1$ attributes only requires the cost to calibrate cuboids with $k$ attributes.

More generally, given an OLAP query that groups by $h$ attributes (so it contains $h$ group-by annotations ${\bf A_h}$), it is executed over a \cjt with $k$ attributes by finding the annotations that differ between the pivot and new query ${\bf A_{h-k}}$, and performing message passing over the associated steiner tree (\Cref{sss:cjtqexec}).  Further, since the calibrated pivot queries span all combinations of $k$ attributes, we simply need to find the pivot query that results in the steiner tree that spans the fewest bags. The optimal pivot query that minimizes the steiner tree could be found in time polynomial in $r$ through dynamic programming, and the algorithm is in \Cref{s:min_steiner}. 

To summarize, calibration of all pivot queries with $k$ attributes costs $O(rn(dm)^k)$, and cost to execute an OLAP query with $h>k$ attributes is $O(s({\bf A_{h-k}})\times\phi)$, where  $s({\bf A_{h-k}})$ is the number of bags in the steiner tree spanning ${\bf A_{h-k}}$, and $\phi$ is the size of the absorption result in $O(nd^{h-1})$, which upper bounds the message size.  

\subsubsection{OLAP Construction Procedure}
Suppose we wish to materialize all cuboids with up to $h$ attributes.  Our complexity analysis shows that there is a space-time tradeoff.   To minimize the time complexity, we calibrate pivot queries with $h-1$ attributes, so that materializing cuboids with $h$ attributes is $O(1)$.  However, calibrating pivot queries with fewer attributes reduces build sizes at the expense of larger steiner trees during cuboid computation. 

\begin{figure}
  \centering
  \includegraphics[width=.8\columnwidth]{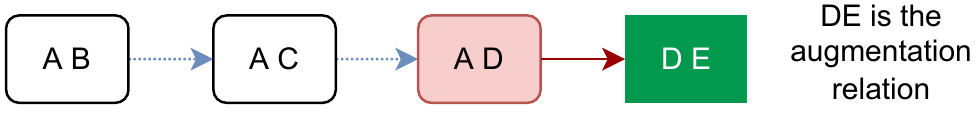}
  \caption{Augmenting the join graph with relation \texttt{DE}. The Steiner Tree is  \texttt{AD} $\rightarrow$ \texttt{DE} with root \texttt{DE} and only requires one message in solid red.}
  \label{fig:augmentation}
\end{figure}

\subsubsection{Comparison With Data Cubes}
From an algebraic perspective,
data cubes exploit the addition operator of the aggregation semi-ring to marginalize out unwanted attributes. 
\cjt further exploits the multiply operator to join relations,
as well as its distributive properties for early marginalization.
In terms of complexity, join materialization alone costs $O(n^r)$ under traditional cube construction, which dominates the total costs of using \cjts.

\subsection{Machine Learning Augmentation}
\label{sec:data_aug}

Data and feature augmentation~\cite{chepurko2020arda} is used to identify datasets to join with an existing training corpus in order to provide more informative features, and is a promising application on top of data lakes and markets~\cite{fernandez2020data,chepurko2020arda,fernandez2018aurum}.   However, the major bottleneck is the cost of joining each augmentation dataset and then retraining the ML model.

The state-of-the-art approach uses factorized learning~\cite{nikolic2018incremental,schleich2016learning,schleich2019layered,curtin2020rk} to avoid join materialization when training models over join graphs.  First, it designs semi-ring structures for common ML models (linear regression~\cite{schleich2016learning}, factorization machines~\cite{schleich2021structure}, k-means~\cite{curtin2020rk}), and then performs early marginalization by pushing the aggregation (training) through the join.   This is equivalent to upward message passing through the join graph.    If we augment with relation $r$, then factorized learning approaches execute message passing through the whole augmented join graph again.  

In contrast, the \cjt allows us to choose any bag $b$ that contains the join keys, construct an edge $b\to r$, and perform message passing using $r$ as the root.  In this setting, the steiner tree is exactly 2 bags, and the rest of the messages in the \cjt can be reused.    For instance, \Cref{fig:augmentation} shows a join graph $\texttt{AB}\to \texttt{AC}\to \texttt{AD}$ that we augment with \texttt{DE}.  The steiner tree is simply \texttt{AD} and \texttt{DE}, and we only need to send one message to compute the updated ML model.  

Although the above is likely the common case, the augmentation relation may have join keys that span multiple bags in the \cjt.  In these cases, the steiner tree spans the bags containing the join keys as well as the new relation.  We discuss details in \Cref{s:augmentation}.

\subsection{Additional Extensions}
\label{ivmjt}

A benefit of the \cjt data structure is that it composes with existing optimizations.  Below we outline two extensions as examples: incremental maintenance and lazy calibration.

\stitle{Incremental Maintenance.}
We may wish to incrementally maintain the \cjt data structure as the base relations are updated.  For instance, stream processing continuously appends records~\cite{abadi2005design}, while counterfactual query explanations~\cite{roy2014formal,wu2013scorpion} delete records (intervention) that satisfy a predicate (explanation) to evaluate their effects on an aggregation result.  We directly apply Factorized-IVM~\cite{nikolic2018incremental} to maintain the \cjt---deltas over the base relations are sent as upward and downward delta messages.  If the delta records are deletions, then the aggregation must be a ring that supports minus operator.

\stitle{Lazy Calibration.}
Calibration is the dominant cost of using \cjts, particularly if we wish to maintain \cjts under updates, because each base relation update will invalidate half of the messages and they need to be re-calibrated. In the spirit of lazy view maintenance~\cite{zhou2007lazy, colby1996algorithms}, we implement lazy calibration.  When a relation is updated, its corresponding bag sends an invalidation message.   A query over the \cjt checks the validity of the messages in its steiner tree, and recalibrates the invalid messages by sending messages from the updated base relation to the invalid bag.    There is an interesting trade-off between the frequency of eager calibration, and recalibration during query processing. The experiment in \Cref{exp:lazycal} shows that lazy calibration can improve performance by $2000\times$ for write-heavy workloads.

\section{Experiments}

\begin{figure}
  \centering
  \includegraphics[width=.75\columnwidth]{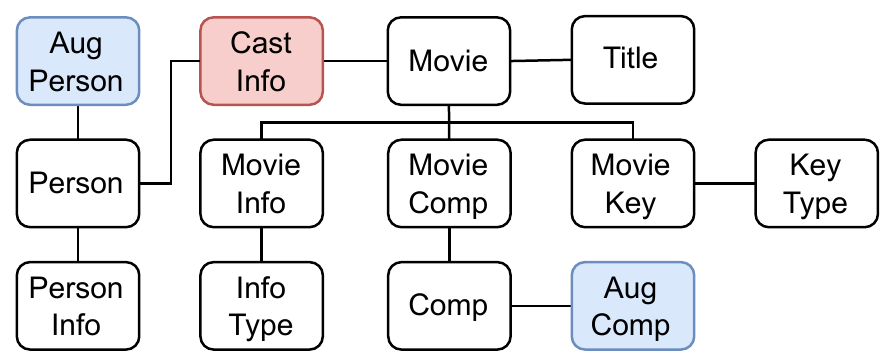}
  \vspace*{-3mm}
  \caption{IMDB schema. \texttt{\red{CastInfo}} is the largest relation. \texttt{\blue{Aug Person}} and \texttt{\blue{Aug Comp}} are augmentation relations.}
  
   \label{imdb_schema}
\end{figure}

Does \cjt really accelerate \analytics including OLAP, query explanation, and ML augmentation applications? 
We have implemented three versions of \cjt: as a custom C++ query engine that uses worst-case optimal join~\cite{gottlob2009generalized}, as a middleware compiler for cloud databases (Redshift), and as a compiler to Pandas DataFrame operations.  We use them to study the above questions.

\subsection{Single-node Custom-engine Experiments}
\label{s:local_exp}
We first study the benefits of worksharing and Factorized-IVM by comparing our custom \cjt engine with \texttt{LMFAO}~\cite{schleich2019layered}, the state-of-the-art factorized query compiler using the wide-table applications in \Cref{s:apps} (OLAP, intervention, ML augmentation). 

\stitle{Setup.} We compare \cjt, \texttt{LMFAO}~\cite{schleich2019layered},  and a variation of our implementation that doesn't perform calibration (\jtivm).  \jtivm uses a similar query execution algorithm as \texttt{LMFAO} (message passing) but, for intervention queries, it also maintains the total count and executes intervention queries using Factorized-IVM~\cite{nikolic2018incremental}.  However, IVM is not applicable to OLAP queries (the aggregation query changes) nor augmentations (the join graph changes). The current implementation of \texttt{LMFAO} only supports queries for a specific application (e.g. linear regression) but not general SPJA queries, so we modified its query compiler to support SPJA queries.  Queries in \texttt{LMFAO} are run one-by-one because we focus on \analytics rather than a batch of queries, and because the current optimizer does not batch SPJA queries (verified with their authors).

We use the IMDB~\cite{imdb} movie dataset (\Cref{imdb_schema}), a popular benchmark dataset in query optimization~\cite{marcus2020bao,leis2015good,marcus2019neo}. Following prior works~\cite{aberger2017emptyheaded,raman2013db2}, we preprocess the dataset to dictionary encode strings into 32-bit unsigned integers. The resulting data is 1.2GB over 11 relations, and the \texttt{Cast Info} relation dominates (${\sim}1GB$).  For space reasons, results on 3 additional datasets (Favorita~\cite{favorita}, Lego~\cite{lego}, and TPC-DS~\cite{nambiar2006making}) can be found in \cite{techreport}.  

We run the systems single threaded on a GCP n1-standard-16 VM, running Debian 10, Xeon 2.20GHz CPU, and 60GB RAM. Because we focus on query processing and for a fair comparison, we exclude the time of reading files from disk and compilation time for \texttt{LMFAO}, but include all the time to build data structures and run queries. All experiments fit and run in memory.

\stitle{Workloads.} 
Our goal is to study the algorithmic benefits of work sharing, so count aggregation queries are sufficient.   Our experiment will calibrate the total count query (count with no grouping attributes) as the pivot query, and execute representative queries for each application class.  For OLAP, we execute two group-by queries 
${\bf Q_1} = \gamma_{COUNT(\cdot), Person}(\Join)$, and
${\bf Q_2} = \gamma_{COUNT(\cdot), Movie} \sigma_{Company}(\Join)$ over the join graph $\Join$.  For interventions, we remove 10 records from the \texttt{Person} (\texttt{Person-10}) or \texttt{MovieKey} (\texttt{MovieKey-10}) relations and refresh the pivot query result.  For augmentation, we create two-attribute augmentation relations where the first attribute is the primary key of \texttt{Person} (\texttt{Aug Person}) or \texttt{Company} (\texttt{Aug Comp}) and the second is a random integer, and refresh the pivot query result.

\begin{figure}
  \centering
  \includegraphics[width=\columnwidth]{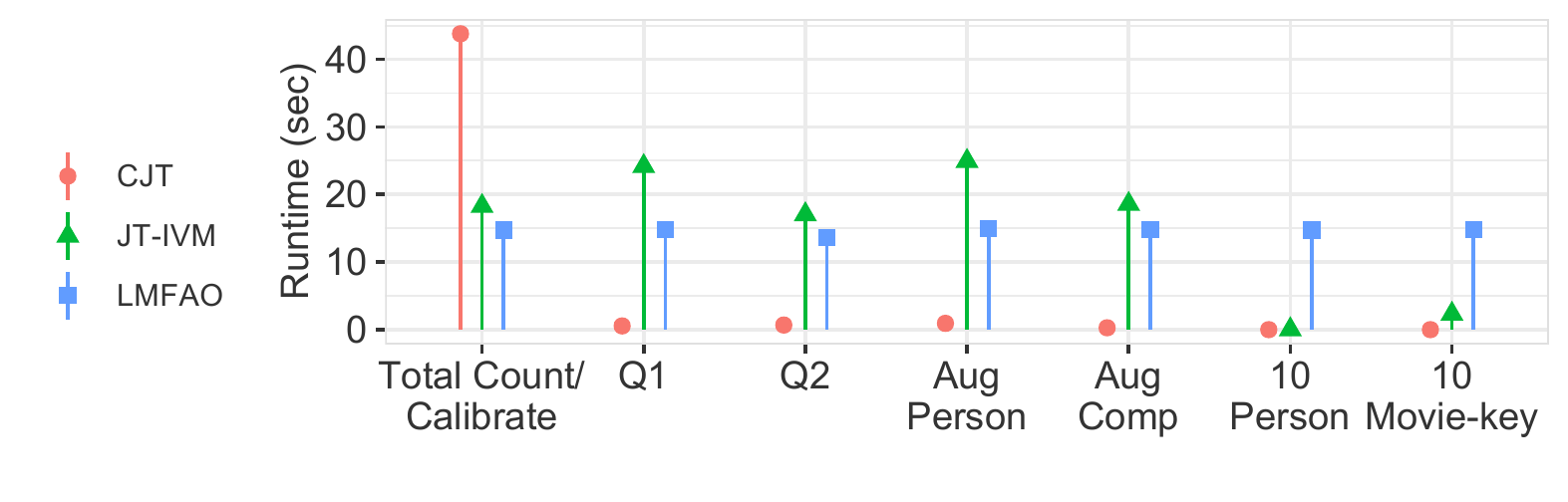}
  \vspace*{-7mm}
  \caption{Run time of different workloads on IMDB dataset.}
  \label{imdb_result}
\end{figure}

\stitle{Results.} 
\Cref{imdb_result} reports the calibration time along with the query times.  
For the calibration overhead of only ${\sim}2\times$ the cost of running the pivot query using \texttt{LMFAO}, \cjt executes OLAP queries  ${\sim}30\times$ faster than \texttt{LMFAO}.  The calibration cost is due to the downward message pass and write overheads, but allows later queries to avoid messages that span the large \texttt{Cast Info} relation.

Similarly, CJTs accelerate augmentation from ${\sim}15s$ (\texttt{LMFAO}) to  ${\sim}1s$ (\cjt); augmenting \texttt{Person} ($18$MB) is $3.1\times$ larger  than \texttt{Company} ($5.8$MB), and so takes ${\sim}4\times$ longer to augment. \jtivm is algorithmically identical to \texttt{LMFAO} for each OLAP query, and is $1.2{\sim}1.7\times$ slower than \texttt{LMFAO} due to implementation differences.

\cjt accelerates intervention queries over \texttt{LMFAO} by ${>}10^5\times$ because the steiner tree is simply the intervened relation/bag, and no messages are passed.
\jtivm naturally out-performs \texttt{LMFAO} as it applies factorized IVM, however it still needs to send messages across the join graph, and is dominated by message sizes (\texttt{Persons-10} is ${\sim}128\times$ faster than \texttt{MovieKeys-10}). This highlights the fact that IVM relies on low fan-out paths to achieve performance.

\subsection{SQL Compiler Cloud Experiments}
\label{exp:cloud}
We now evaluate \cjt (the compiler middleware) on AWS Redshift using SPJA queries over wide-tables.  \cjt is initialized with the join graph, and translates the pivot query into a set of \texttt{CREATE TABLE} statements---after picking a random root, each message during upward and downward message passing generates a query with the corresponding join, filter, and aggregation logic.  Delta queries over the \cjt are similarly translated.  

\stitle{Setup.} 
We use three datasets: TPC-H (SF=10), TPC-DS (SF=1), and \texttt{synthetic}.   Following prior \jt work~\cite{xirogiannopoulos2019memory}, \texttt{synthetic} contains $r\in[2,8]$ relations with a chain schema: 
$$R(A_1, A_2), R(A_2,A_3),\ldots, R(A_r, A_{r+1}).$$ 
We vary the fanout $f$ between adjacent relations (low=2, mid=5, high=10), and the attribute domain size $d$. For each value of $A_i$ in $R(A_i, A_{i+1})$, we assign $f$ unique values to $A_{i+1}$ with fanout f being implemented by, for each value in $A_i$, assigning f sequential values to $A_{i+1}$, such that the $n^{th}$ value is $n\%d$. Thus, the fanout $f$ is in both directions.  We vary the fanout (and domain) and keep the total join size $d\times f^8$ fixed to be $10^9$. The domain sizes $d$ for different fanouts are $d_{low}=3906250$, $d_{mid}=2560$, and $d_{high}=10$.

We used an AWS redshift cluster (one dc2.large node, 2 vCPU, 15 GB memory, 0.16TB SSD, 0.60 GB/s I/O). All experiments warm the cache by pre-executing queries until the runtime stabilizes.

\begin{figure}
  \centering
  \includegraphics[width=0.65\columnwidth]{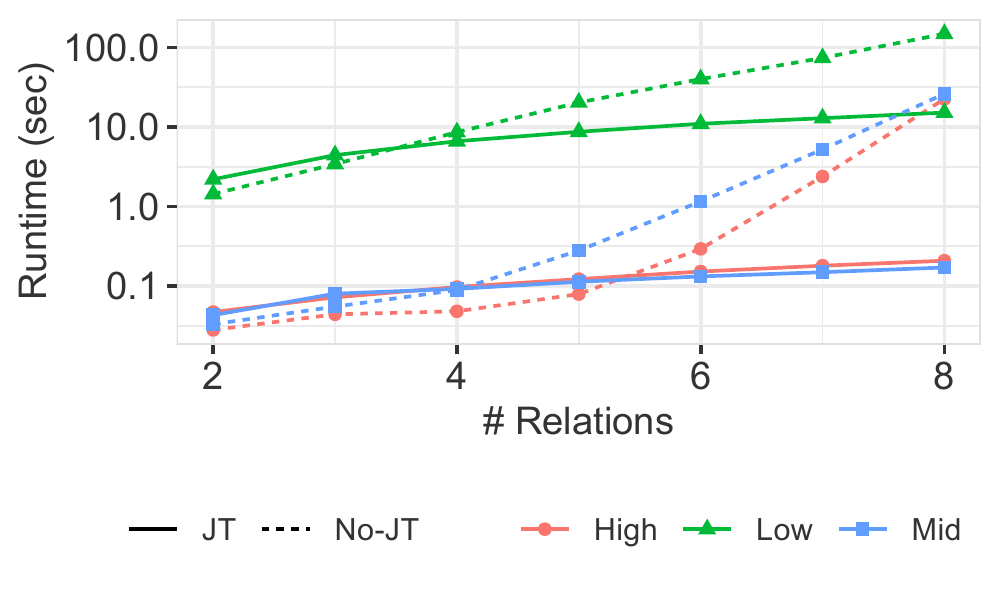}
  \vspace*{-5mm}
  \caption{Run time of total count query with (\texttt{JT})/without message passing (\texttt{No-JT}) in seconds (log scale). High, Mid, and Low are for different fanouts.}
  \label{fig:exp_msg} 
\end{figure}

\subsubsection{Message Passing Costs}
We first evaluate the benefits of message passing (but not calibration) in cloud settings.  The compiler generates \texttt{CREATE VIEW} statements, so that messages are {\it not} materialized.  We execute the total count query as a large join-aggregation query (\texttt{No-JT}) or as an upward message passing (\texttt{JT}).

\Cref{fig:exp_msg} varies the number of relations (x-axis) and fanout (line marker).  Message passing reduces the runtimes from exponential to linear due to early marginalization, but incurs a small overhead to perform marginalization when there are few relations.  Low fan-out has the largest runtime because we fix the total join size and hence the low fan-out has the largest domain size. Note that the x-axis is also interpretable as the steiner tree size.

\begin{figure}[h]
  \includegraphics[width=0.6\columnwidth]{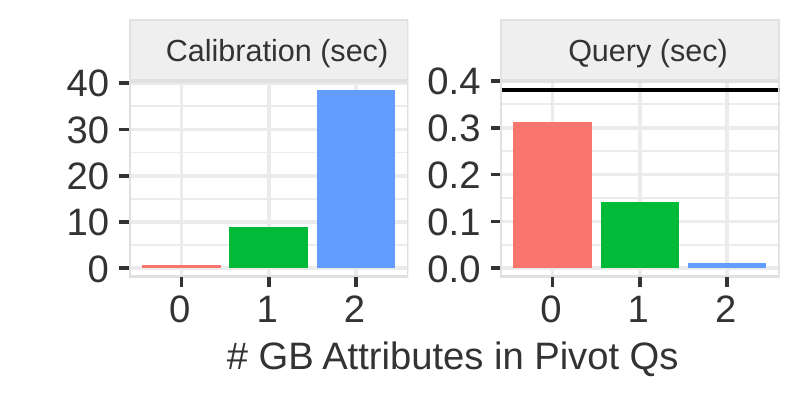}
  \vspace*{-5mm}
  \caption{We vary the dimensionality of the calibrated pivot queries $k \in \{0,1,2\}$ and measure calibration runtime and impact on 4-attribute OLAP queries.  Horizontal line represents the average runtime with \jt.
  }  
  \label{exp:cube_calibration}
\end{figure}

\subsubsection{Cubes in the Cloud} 
\cjts help developers build data cubes to explicitly trade-off build costs and query performance.  To evaluate this, we use the synthetic dataset with $f=10$ (high) fanout and $r=8$ relations, and calibrate all cuboids with $k\in[1,3]$ grouping attributes.  For each $k$, we use the cuboids to execute 100 random OLAP queries with 4 grouping attributes.

The results are in \Cref{exp:cube_calibration}.  Although calibration cost increases exponentially (as expected), message passing is still significantly faster than naive query execution: computing {\it all} 2-attribute cuboids (through calibration of all 1 group-by attribute Pivot Qs in $8.8$s) is substantially faster than naively computing a {\it single} 0-attribute cuboid  ($22.6$s for \texttt{No-JT} in \Cref{fig:exp_msg}). At the same time, increasing the dimensionality of the cuboids ($k$) significantly reduces the query runtimes ($2.71\times$ speedup for $k=1$, and $33.73\times$ speedup for $k=2$) due to the smaller steiner tree.

\begin{figure}
  \centering
      \includegraphics [width=0.95\columnwidth] {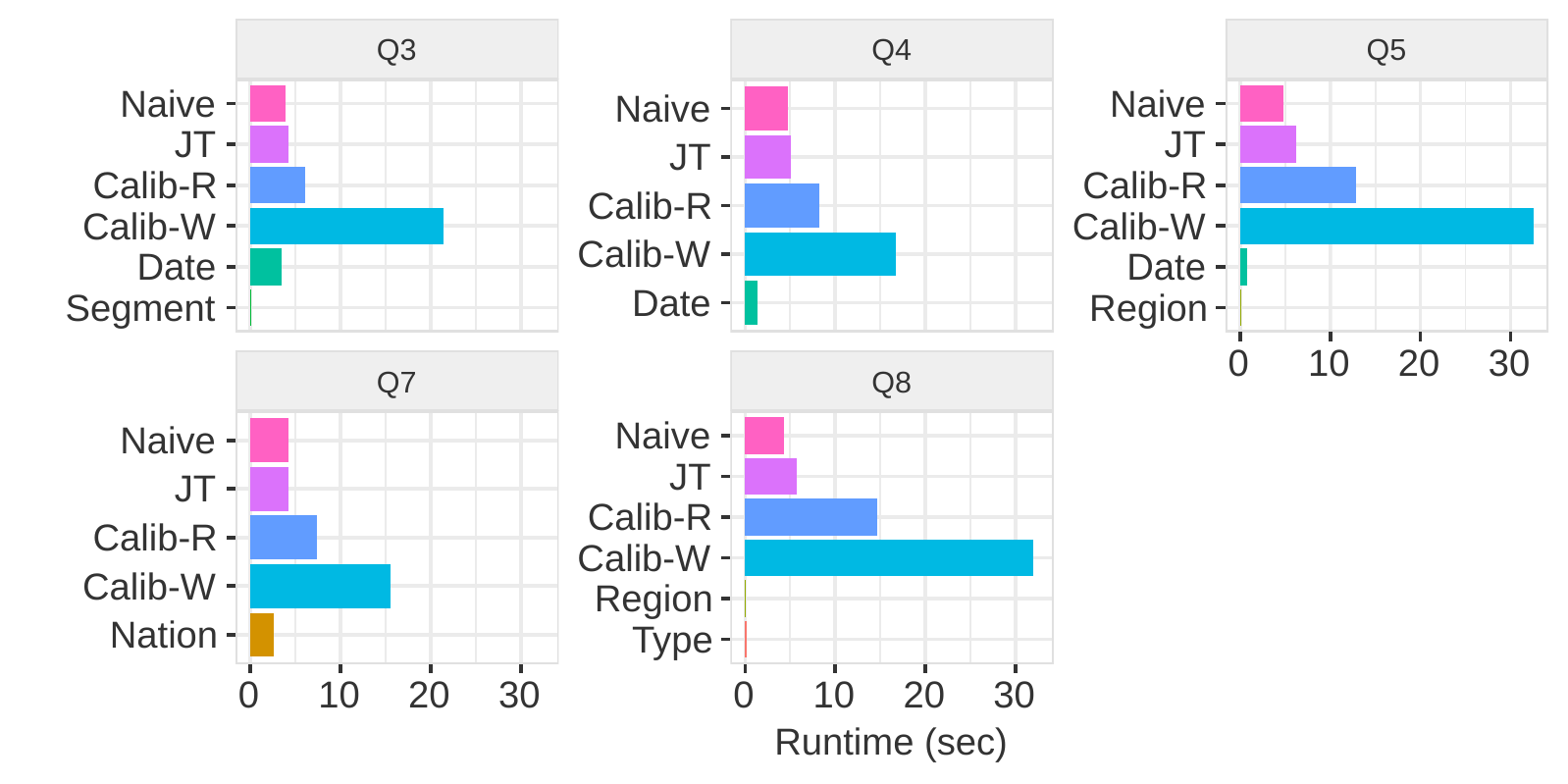}
      \vspace*{-5mm}
  \caption{Run time for TPC-H queries. \texttt{Naive} is for original queries without message passing. \texttt{Calib-R} is for computing all messages without materializing them. \texttt{Calib-W} is for computing and materializing messages.}
  \label{fig:tpch_runtime}
\end{figure}

\subsubsection{TPC-H}
TPC-H queries contain parameterized queries that can benefit from \cjt.   We use a subset of TPC-H queries (Q3-5,7,8) that can be re-written as acyclic SPJA queries (see \Cref{s:tpchqueries}). For each parameterized query, we calibrate the pivot query with random parameter values, and then vary the parameters one at a time as delta queries.

\Cref{fig:tpch_runtime} reports the calibration and query execution costs.  \texttt{Naive} simply runs the query on Redshift. For \cjt, we report calibration execution cost (\texttt{Calib-R}) separately from the calibration materialization cost (\texttt{Calib-W}), since writes on Redshift are particularly expensive. We also attempted to compare \cjt against previous factorized systems~\cite{schleich2019layered,olteanu2015size}  like \texttt{LMFAO}, but they are not distributed. Therefore, we report \jt, which uses message passing and is algorithmically similar to \texttt{LMFAO} for individual queries.
The remaining bars correspond to varying corresponding parameter value.

Calibration (\texttt{Calib-W}) takes $4{\sim}7\times$ longer than \texttt{Naive}.  As expected, upward and downward message passing alone is ${\sim}2\times$ slower (\texttt{Calib-R}), and the rest is dominated by high write overheads;   Q8 groups by 2 attributes, so its message sizes are ${\sim}2\times$ larger, and $4\times$ slower overall.  
\jt is about $2\times$ faster than calibration alone (\texttt{Calib-R}). This is because, for a {\it single} query, \jt is equivalent to upward message passing.  However, we can see that factorized execution alone does not execute the TPC-H queries tangibly faster than \texttt{Naive}.    This is because factorization is optimized for many-to-many joins, however TPC-H is dominated by one-to-many joins, so the benefits are minimal.

\begin{figure}
  \centering
  \begin{subfigure}[c]{0.25\textwidth}
     \centering
     \includegraphics[width=\textwidth]{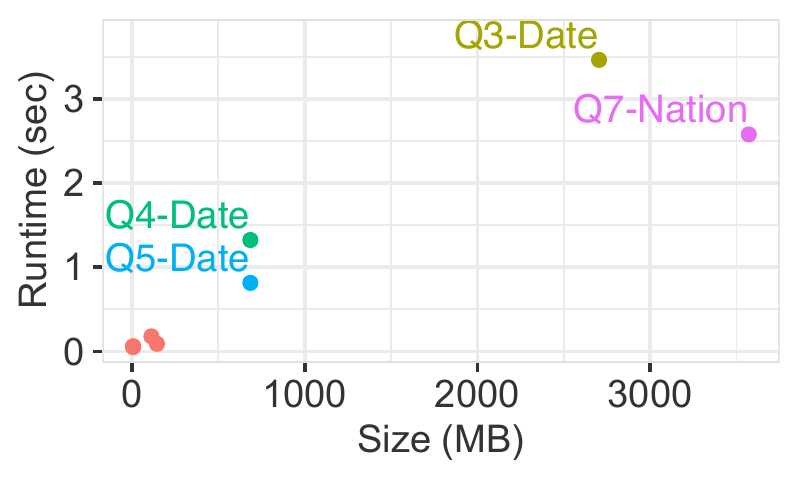}
     \vspace*{-5mm}
     \caption{}
     \label{fig:tpch_runtime_size}
 \end{subfigure}
 \hfill
  \begin{subfigure}[c]{0.18\textwidth}
     \centering
     \includegraphics[width=\textwidth]{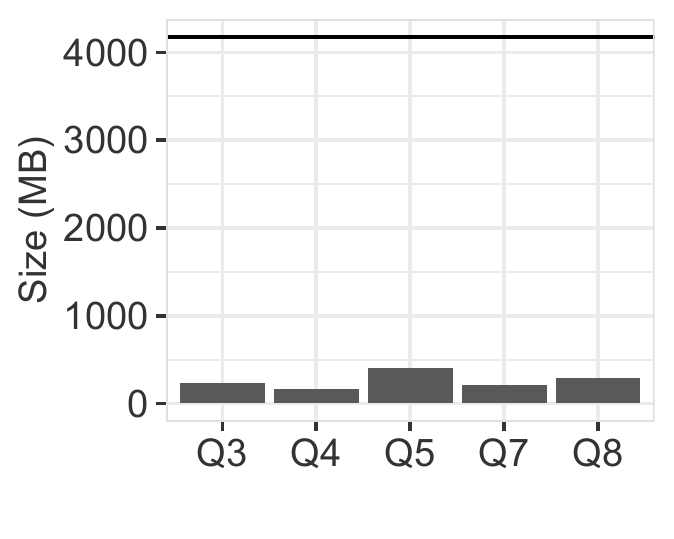}
     \vspace*{-5mm}
     \caption{}
     \label{fig:tpch_size}
 \end{subfigure}
 \hfill
\vspace*{-3mm}
  \caption{(a) Size of annotated bag (by predicate) vs query runtime. (b) Total message size overhead of \cjt. Horizontal line is the TPC-H database size (4172 MB).}
\end{figure}

In contrast, \cjt accelerates TPC-H queries by nearly $1000\times$ over \texttt{Naive} for multiple parameters including Segment, Region and Type. The speedup for other parameters is less significant.  Naturally, the speedup depends linearly on the size of the bag that contains the parameterized attribute (\Cref{fig:tpch_runtime_size}). We note that the space overhead for calibration is only ${<}500 MB$ compared to the original database size $4172 MB$ (\Cref{fig:tpch_size}). This is because messages are aggregated results from these relations.

\begin{figure}
  \centering
      \includegraphics [width=0.3\textwidth] {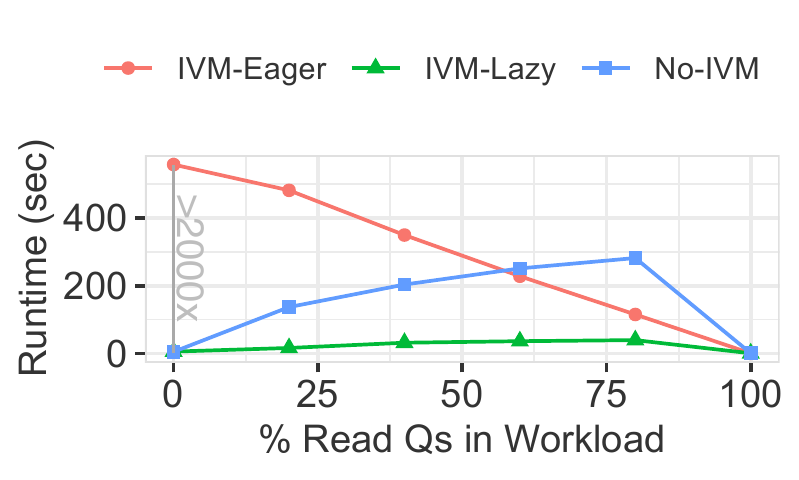}
      \vspace*{-5mm}
  \caption{Runtime for read-write workload (100 Qs).}
  \label{fig:lazycal}
\end{figure}

\subsubsection{Lazy Calibration}
\label{exp:lazycal}
We now use TPC-DS (SF=1) on a read-write workload (100 queries) to showcase the trade-offs of lazy calibration (\texttt{Lazy-IVM}) compared to lazy re-evaluation w/out IVM (\texttt{No-IVM}) or eager calibration (\texttt{Eager-IVM}).    The database contains 9 relations. Each write query inserts a random row into the \texttt{Store Sales} relation (largest relation, 181MB), and each read query randomly chooses one of the other relations ($1-42$MB) and computes count grouped by its primary key.

\Cref{fig:lazycal} reports total workload cost with varying read-write mixtures.  \texttt{Eager-IVM} linearly improves as it contains fewer writes, while both lazy approaches increase in cost with read \%. \texttt{No-IVM} is more expensive due to full message sizes rather than deltas.    The costs are near-zero at the extremes when there are no reads (no message passing) or no writes (no maintenance needed).
Eager calibration w/out IVM is not plotted because it is ${>}1$hr at 20\% writes.

\subsection{Pandas Dataframe Compiler}

Python and Pandas is one of the dominant programming environments for ML.
Thus we evaluate the Pandas~\cite{mckinney2010data} dataframe compiler for an ML data augmentation task using linear regression~\cite{schleich2016learning}. The compiler takes the join graph and relations (as Pandas dataframes) as input, and then translates calibration, model training, model evaluation and data augmentation into Pandas merge (join), group-by aggregation and array operations. We also report the model training time using \texttt{LMFAO}~\cite{schleich2019layered} for reference.

\begin{figure}
  \centering
  \includegraphics[width=.6\columnwidth]{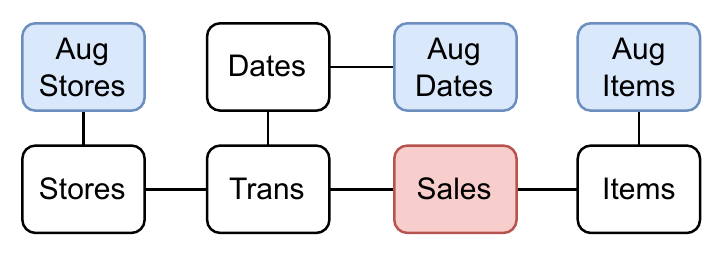}
  \vspace*{-3mm}
  \caption{Favorita schema. \texttt{\red{Sales}} is the largest relation. \texttt{\blue{Aug Stores}}, \texttt{\blue{Aug Dates}} and \texttt{\blue{Aug Items}} are augmentation relations.}
   \label{fav_schema}
\end{figure}

\stitle{Setup.} We use the Favorita~\cite{favorita} dataset of purchasing and sales forecasts, used in prior factorized learning works~\cite{schleich2016learning, schleich2019layered} (schema in \Cref{fav_schema}). \texttt{Sales} is the largest relation (241MB), and the others are $<2MB$.  The model uses (\texttt{Sales}.unit\_sales, \texttt{Stores}.type, \texttt{Items}.perishable) as features, and \texttt{Trans}.transactions (number of transactions for each store, date) as the target Y variable. 

To simulate a data lake with augmentation data of varying effectiveness, we generate synthetic data to join with (augment) \texttt{Dates}, \texttt{Stores}, and \texttt{Items}.  For each relation $R$ (e.g., \texttt{Dates}), we first generate a predictive feature $\hat{Y}$ as the average of Y grouped by $R$'s primary key. Then we create 10 augmentation relations with schema \texttt{(key,val)}, where \texttt{key} is the join attribute and \texttt{val} varies in correlation $\hat{Y}$~\cite{kaiser1962sample}.  The correlation coefficient $\phi$ is drawn from the inverse exponential distribution: min(1, 1/Exp(10)), and the values are the weighed average between $\hat{Y}$ and a random variable, weighed by $\phi$. 

We individually evaluate the model for each of the 30 augmentation relations, and measure the cumulative runtimes and model accuracy (R2).  We used the GCP machine in \Cref{s:local_exp}.

\begin{figure}
  \begin{subfigure}[t]{0.49\columnwidth}
  \centering
    \includegraphics[width=\textwidth]{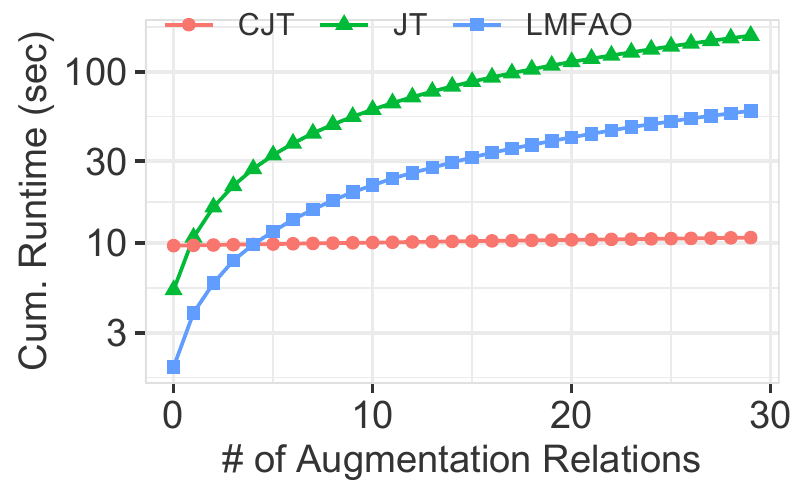}
    \caption{Cumulative runtime (log)}
     \label{exp:aug_time}
     \end{subfigure}
     \hfill
    \begin{subfigure}[t]{0.49\columnwidth}
    \centering
    \includegraphics[width=\textwidth]{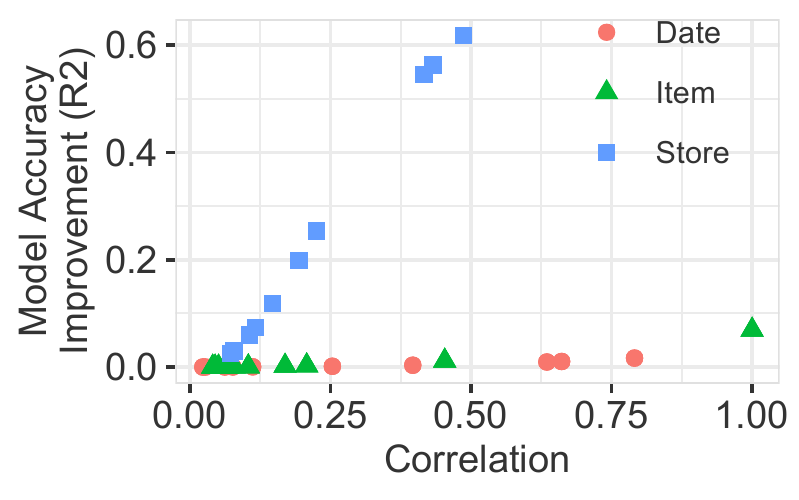}
    \caption{Accuracy (R2) improvements.}
    \label{exp:aug_r2}
     \end{subfigure}
     \vspace*{-3mm}
  \caption{Augmentation run time and model performance.}
\end{figure}

\stitle{Results.} 
\Cref{exp:aug_time} reports the cumulative costs to augment and retrain the model.
 Factorized learning (\jt) takes ${>}1.5$ min, while \cjt takes ${\sim}10$ sec: calibration dominates the cost, and is ${\sim}2\times$ the cost of training a single model because join costs dominate. After calibration, \cjts can simply evaluate all 30 augmentations in ${<}1$ sec.  
\texttt{LMFAO} takes ${\sim}2\times$ less time than \jt for model training due to implementation differences (see \Cref{s:local_exp}). Even with the \cjt build cost, \cjt is ${\sim}6\times$ faster than \texttt{LMFAO} after 30 augmentations.
\Cref{exp:aug_r2} reports the accuracy improvement above the baseline ($0.031$) after each augmentation, and we see a wide discrepancy between good and bad augmentations ($+0$ to $+0.61$).  

\section{Related Work}\label{s:related}

\stitle{Wide-Table and Factorized Join.} 
The wide-table in-memory columnar engine~\cite{li2014widetable} shows that denormalized query execution can speed TPC-H queries by up to $10\times$.  However, it is limited to in-memory relations, and denormalized relations can be too large to fit in the memory. Kumar et al.~\cite{kumar2015learning} use user-defined aggregate functions to push ML computations over factorized join~\cite{olteanu2015size} and show improvement in run time and storage. However, their algorithm only improves the space but not time complexity, and performs {\it worse} than the naive approach when the data fits in the memory.

\stitle{Early Marginalization.} Early Marginalization was first introduced by Gupta et al.~\cite{gupta1995aggregate} as a generalized projection for simple e.g., count, sum, max queries. It was extended by factorized databases to compactly store relational tables~\cite{olteanu2015size} and quickly execute semi-ring aggregation queries~\cite{schleich2019layered,joglekar2015aggregations}. Abo et al.~\cite{abo2016faq} generalize early marginalization and establish the equivalence between early marginalization and variable elimination in Probabilistic Graphical Models~\cite{koller2009probabilistic}. 
However, prior works don't exploit work sharing for \analytics. \texttt{LMFAO}~\cite{schleich2019layered} shares work within a query batch but not between batches; \texttt{IFAQ}~\cite{shaikhha2020multi} iteratively performs batch gradient descent, but the key aggregation (covariance) is computed before iteration; and \texttt{FIVM}~\cite{nikolic2018incremental} focuses on a single query.

\stitle{Calibrated Junction Tree.} Calibration Junction Tree is first proposed by Shafer and Shenoy~\cite{shafer1990probability} to compute inference over probabilistic graphical models. Calibration Junction Tree has been widely used across engineering~\cite{zhu2015junction,ramirez2009fault}, ML~\cite{braun2016lifted,deng2014large}, and medicine~\cite{pineda2015novel,lauritzen2003graphical}, but they are limited to probabilistic tables. Yannakakis's algorithm~\cite{yannakakis1981algorithms} applies two-pass semi-join reduction to relations, but is limited to 0/1 semi-ring.
We generalize the Calibration Junction Tree to semi-ring aggregation  and extend it to support SPJA queries. 

\stitle{Constant Delay Enumeration.} Efficient enumeration of conjunctive queries (w/boolean semi-ring) is widely studied. For free-connex queries~\cite{bagan2007acyclic}, we can build \jt in $O(n)$ to enumerate results in $O(1)$; under updates, we can build and maintain simple generalized join trees (\texttt{SGJT})~\cite{idris2017dynamic} in $O(n)$.  Q-hierarchical queries~\cite{berkholz2017answering, kara2020trade} reduce maintenance to $O(1)$ by building view trees (\texttt{VT}). The main idea in both is to maintain materialized views that exploit the query structure and create short-cuts for enumeration.  
\texttt{SGJT} and \texttt{VT} can be expressed by \cjt using empty bags. For \texttt{SGJT} (resp. \texttt{VT}), each leaf node (resp. atom node) is a bag containing the relation, and each interior node (resp. view node) is an empty bag with the same schema; \cjt calibration materializes the views and performs semi-join reduction for \texttt{SGJT} and \texttt{VT}. Note that \cjt is more expressive and expresses optimizations~\cite{kara2018counting} for cyclic queries as well (\Cref{latency_cjt}).

\stitle{Materialized View.} Previous materialized view works ~\cite{asgharzadeh2009exact,li2005formal} focus on the optimization of view selections given workloads; these works have significant overhead and rely on heuristics to solve the intractable optimization. \cjt is lightweight with the same complexity of pivot query. \cjt is closely related to and generalizes data cube~\cite{gray1997data} as it materializes views of the full join result; as shown in \Cref{exp:cloud}, \cjt is much more efficient than data cube. Recent higher-order IVM~\cite{ahmad2012dbtoaster} and Factorized-IVM~\cite{nikolic2018incremental} exploit the semi-ring structure to optimize IVM, which also benefits \cjt (\Cref{s:local_exp}).

\stitle{Machine Learning Systems.} Traditional ML systems only accept a single relation as input~\cite{ghoting2011systemml,hellerstein2012madlib} and require the materialization of wide-table. Structure-aware ML systems ~\cite{schleich2016learning,schleich2019layered} apply early marginalization to train model over wide-table efficiently. \cjt contributes to Machine Learning Systems in two ways: (1). \cjt supports efficient ML augmentation. (2). \cjt shares computations between tree-based models like Random Forest by re-using messages from low-dimensional cuboids discussed in \Cref{app:olap}.

\section{Conclusions}

This work presented \cjt, a novel data structure based on calibration to materialize and manages messages for \analytics.   Delta queries are able to re-use messages to accelerate execution by orders of magnitude as compared to factorized query execution.  To do so, we developed an annotation-based analysis to identify reusable messages.   We showed how \cjts accelerate OLAP workloads, query explanation, streaming data, and data augmentation for ML; and benefits from incremental view maintenance and lazy calibration.  We evaluated three implementations of \cjt---a single-node query engine and middleware compilers to SQL and Pandas dataframe operations---on local and cloud databases.  Single-node engine accelerates OLAP queries by ${\sim}30\times$ and intervention queries by over $10^5\times$ over the state-of-the-art factorized engine LMFAO; cloud compiler accelerates TPC-H queries on Redshift by up to $10^3\times$ compared to factorized execution; and Pandas compiler accelerates data augmentation for ML by ${>}100\times$.

\balance
\bibliographystyle{abbrv}
\bibliography{paper}

\begin{thebibliography}{10}

\bibitem{imdb}
Imdb.
\newblock \url{https://www.imdb.com/interfaces/}, 05 2013.

\bibitem{favorita}
Corporación favorita grocery sales forecasting.
\newblock \url{https://www.kaggle.com/c/favorita-grocery-sales-forecasting}, 10
  2017.

\bibitem{lego}
Lego database.
\newblock \url{https://www.kaggle.com/rtatman/lego-database}, 07 2017.

\bibitem{abadi2005design}
D.~J. Abadi, Y.~Ahmad, M.~Balazinska, U.~Cetintemel, M.~Cherniack, J.-H. Hwang,
  W.~Lindner, A.~Maskey, A.~Rasin, E.~Ryvkina, et~al.
\newblock The design of the borealis stream processing engine.
\newblock In {\em Cidr}, volume~5, pages 277--289, 2005.

\bibitem{aberger2017emptyheaded}
C.~R. Aberger, A.~Lamb, S.~Tu, A.~N{\"o}tzli, K.~Olukotun, and C.~R{\'e}.
\newblock Emptyheaded: A relational engine for graph processing.
\newblock {\em ACM Transactions on Database Systems (TODS)}, 42(4):1--44, 2017.

\bibitem{abo2016faq}
M.~Abo~Khamis, H.~Q. Ngo, and A.~Rudra.
\newblock Faq: questions asked frequently.
\newblock In {\em Proceedings of the 35th ACM SIGMOD-SIGACT-SIGAI Symposium on
  Principles of Database Systems}, pages 13--28, 2016.

\bibitem{ahmad2012dbtoaster}
Y.~Ahmad, O.~Kennedy, C.~Koch, and M.~Nikolic.
\newblock Dbtoaster: Higher-order delta processing for dynamic, frequently
  fresh views.
\newblock {\em arXiv preprint arXiv:1207.0137}, 2012.

\bibitem{albutiu2012massively}
M.-C. Albutiu, A.~Kemper, and T.~Neumann.
\newblock Massively parallel sort-merge joins in main memory multi-core
  database systems.
\newblock {\em arXiv preprint arXiv:1207.0145}, 2012.

\bibitem{alon1997finding}
N.~Alon, R.~Yuster, and U.~Zwick.
\newblock Finding and counting given length cycles.
\newblock {\em Algorithmica}, 17(3):209--223, 1997.

\bibitem{arun2018hypertree}
A.~S. Arun, S.~V.~M. Jayaraman, C.~R{\'e}, and A.~Rudra.
\newblock Hypertree decompositions revisited for pgms.
\newblock {\em arXiv preprint arXiv:1804.01640}, 2018.

\bibitem{asgharzadeh2009exact}
Z.~Asgharzadeh~Talebi, R.~Chirkova, and Y.~Fathi.
\newblock Exact and inexact methods for solving the problem of view selection
  for aggregate queries.
\newblock {\em International Journal of Business Intelligence and Data Mining},
  4(3-4):391--415, 2009.

\bibitem{atserias2008size}
A.~Atserias, M.~Grohe, and D.~Marx.
\newblock Size bounds and query plans for relational joins.
\newblock In {\em 2008 49th Annual IEEE Symposium on Foundations of Computer
  Science}, pages 739--748. IEEE, 2008.

\bibitem{bagan2007acyclic}
G.~Bagan, A.~Durand, and E.~Grandjean.
\newblock On acyclic conjunctive queries and constant delay enumeration.
\newblock In {\em International Workshop on Computer Science Logic}, pages
  208--222. Springer, 2007.

\bibitem{berkholz2017answering}
C.~Berkholz, J.~Keppeler, and N.~Schweikardt.
\newblock Answering conjunctive queries under updates.
\newblock In {\em proceedings of the 36th ACM SIGMOD-SIGACT-SIGAI symposium on
  Principles of database systems}, pages 303--318, 2017.

\bibitem{braun2016lifted}
T.~Braun and R.~M{\"o}ller.
\newblock Lifted junction tree algorithm.
\newblock In {\em Joint German/Austrian Conference on Artificial Intelligence
  (K{\"u}nstliche Intelligenz)}, pages 30--42. Springer, 2016.

\bibitem{widetable3}
C.~Carroll.
\newblock Modeling marketing attribution.
\newblock \url{https://blog.getdbt.com/modeling-marketing-attribution/}, 2020.

\bibitem{chepurko2020arda}
N.~Chepurko, R.~Marcus, E.~Zgraggen, R.~C. Fernandez, T.~Kraska, and D.~Karger.
\newblock Arda: automatic relational data augmentation for machine learning.
\newblock {\em arXiv preprint arXiv:2003.09758}, 2020.

\bibitem{joinshard}
@codinglay.
\newblock Is it just me or are joins the most confusing concept to understand
  in sql???
\newblock \url{https://twitter.com/codinglay/status/1501594773685166083}, 2022.

\bibitem{colby1996algorithms}
L.~S. Colby, T.~Griffin, L.~Libkin, I.~S. Mumick, and H.~Trickey.
\newblock Algorithms for deferred view maintenance.
\newblock In {\em Proceedings of the 1996 ACM SIGMOD international conference
  on Management of data}, pages 469--480, 1996.

\bibitem{cozman2000generalizing}
F.~G. Cozman et~al.
\newblock Generalizing variable elimination in bayesian networks.
\newblock In {\em Workshop on probabilistic reasoning in artificial
  intelligence}, pages 27--32. Citeseer, 2000.

\bibitem{curtin2020rk}
R.~Curtin, B.~Moseley, H.~Ngo, X.~Nguyen, D.~Olteanu, and M.~Schleich.
\newblock Rk-means: Fast clustering for relational data.
\newblock In {\em International Conference on Artificial Intelligence and
  Statistics}, pages 2742--2752. PMLR, 2020.

\bibitem{dehne2002parallelizing}
F.~Dehne, T.~Eavis, S.~Hambrusch, and A.~Rau-Chaplin.
\newblock Parallelizing the data cube.
\newblock {\em Distributed and Parallel Databases}, 11(2):181--201, 2002.

\bibitem{deng2014large}
J.~Deng, N.~Ding, Y.~Jia, A.~Frome, K.~Murphy, S.~Bengio, Y.~Li, H.~Neven, and
  H.~Adam.
\newblock Large-scale object classification using label relation graphs.
\newblock In {\em European conference on computer vision}, pages 48--64.
  Springer, 2014.

\bibitem{dong2022table}
Y.~Dong and M.~Oyamada.
\newblock Table enrichment system for machine learning.
\newblock {\em arXiv preprint arXiv:2204.08235}, 2022.

\bibitem{eden2017approximately}
T.~Eden, A.~Levi, D.~Ron, and C.~Seshadhri.
\newblock Approximately counting triangles in sublinear time.
\newblock {\em SIAM Journal on Computing}, 46(5):1603--1646, 2017.

\bibitem{fernandez2018aurum}
R.~C. Fernandez, Z.~Abedjan, F.~Koko, G.~Yuan, S.~Madden, and M.~Stonebraker.
\newblock Aurum: A data discovery system.
\newblock In {\em 2018 IEEE 34th International Conference on Data Engineering
  (ICDE)}, pages 1001--1012. IEEE, 2018.

\bibitem{fernandez2020data}
R.~C. Fernandez, P.~Subramaniam, and M.~J. Franklin.
\newblock Data market platforms: Trading data assets to solve data problems.
\newblock {\em arXiv preprint arXiv:2002.01047}, 2020.

\bibitem{fischl2018general}
W.~Fischl, G.~Gottlob, and R.~Pichler.
\newblock General and fractional hypertree decompositions: Hard and easy cases.
\newblock In {\em Proceedings of the 37th ACM SIGMOD-SIGACT-SIGAI Symposium on
  Principles of Database Systems}, pages 17--32, 2018.

\bibitem{fisher2012exploratory}
D.~Fisher, S.~M. Drucker, and A.~C. K{\"o}nig.
\newblock Exploratory visualization involving incremental, approximate database
  queries and uncertainty.
\newblock {\em IEEE computer graphics and applications}, 32(4):55--62, 2012.

\bibitem{freitag2020adopting}
M.~Freitag, M.~Bandle, T.~Schmidt, A.~Kemper, and T.~Neumann.
\newblock Adopting worst-case optimal joins in relational database systems.
\newblock {\em Proceedings of the VLDB Endowment}, 13(12):1891--1904, 2020.

\bibitem{ghoting2011systemml}
A.~Ghoting, R.~Krishnamurthy, E.~Pednault, B.~Reinwald, V.~Sindhwani,
  S.~Tatikonda, Y.~Tian, and S.~Vaithyanathan.
\newblock Systemml: Declarative machine learning on mapreduce.
\newblock In {\em 2011 IEEE 27th International Conference on Data Engineering},
  pages 231--242. IEEE, 2011.

\bibitem{gonzalez2010google}
H.~Gonzalez, A.~Y. Halevy, C.~S. Jensen, A.~Langen, J.~Madhavan, R.~Shapley,
  W.~Shen, and J.~Goldberg-Kidon.
\newblock Google fusion tables: web-centered data management and collaboration.
\newblock In {\em Proceedings of the 2010 ACM SIGMOD International Conference
  on Management of data}, pages 1061--1066, 2010.

\bibitem{gottlob2009generalized}
G.~Gottlob, Z.~Mikl{\'o}s, and T.~Schwentick.
\newblock Generalized hypertree decompositions: Np-hardness and tractable
  variants.
\newblock {\em Journal of the ACM (JACM)}, 56(6):1--32, 2009.

\bibitem{gray1997data}
J.~Gray, S.~Chaudhuri, A.~Bosworth, A.~Layman, D.~Reichart, M.~Venkatrao,
  F.~Pellow, and H.~Pirahesh.
\newblock Data cube: A relational aggregation operator generalizing group-by,
  cross-tab, and sub-totals.
\newblock {\em Data mining and knowledge discovery}, 1(1):29--53, 1997.

\bibitem{green2007provenance}
T.~J. Green, G.~Karvounarakis, and V.~Tannen.
\newblock Provenance semirings.
\newblock In {\em Proceedings of the twenty-sixth ACM SIGMOD-SIGACT-SIGART
  symposium on Principles of database systems}, pages 31--40, 2007.

\bibitem{gupta1995aggregate}
A.~Gupta, V.~Harinarayan, and D.~Quass.
\newblock Aggregate-query processing in data warehousing environments.
\newblock 1995.

\bibitem{han1998selective}
J.~Han, N.~Stefanovic, and K.~Koperski.
\newblock Selective materialization: An efficient method for spatial data cube
  construction.
\newblock In {\em Pacific-Asia conference on knowledge discovery and data
  mining}, pages 144--158. Springer, 1998.

\bibitem{hellerstein2012madlib}
J.~Hellerstein, C.~R{\'e}, F.~Schoppmann, D.~Z. Wang, E.~Fratkin, A.~Gorajek,
  K.~S. Ng, C.~Welton, X.~Feng, K.~Li, et~al.
\newblock The madlib analytics library or mad skills, the sql.
\newblock {\em arXiv preprint arXiv:1208.4165}, 2012.

\bibitem{techreport}
Z.~Huang and E.~Wu.
\newblock Calibration: A simple trick for wide-table delta analytics (technical
  report).
\newblock
  \url{https://anonymous.4open.science/r/CJT-2DF1/tech_report/Calibrated_Junction_Hypertree.pdf},
  2022.

\bibitem{idris2017dynamic}
M.~Idris, M.~Ugarte, and S.~Vansummeren.
\newblock The dynamic yannakakis algorithm: Compact and efficient query
  processing under updates.
\newblock In {\em Proceedings of the 2017 ACM International Conference on
  Management of Data}, pages 1259--1274, 2017.

\bibitem{joglekar2015aggregations}
M.~Joglekar, R.~Puttagunta, and C.~R{\'e}.
\newblock Aggregations over generalized hypertree decompositions.
\newblock {\em arXiv preprint arXiv:1508.07532}, 2015.

\bibitem{joglekar2018s}
M.~Joglekar and C.~R{\'e}.
\newblock It’s all a matter of degree.
\newblock {\em Theory of Computing Systems}, 62(4):810--853, 2018.

\bibitem{joglekar2016ajar}
M.~R. Joglekar, R.~Puttagunta, and C.~R{\'e}.
\newblock Ajar: Aggregations and joins over annotated relations.
\newblock In {\em Proceedings of the 35th ACM SIGMOD-SIGACT-SIGAI Symposium on
  Principles of Database Systems}, pages 91--106, 2016.

\bibitem{kaiser1962sample}
H.~F. Kaiser and K.~Dickman.
\newblock Sample and population score matrices and sample correlation matrices
  from an arbitrary population correlation matrix.
\newblock {\em Psychometrika}, 27(2):179--182, 1962.

\bibitem{kara2018counting}
A.~Kara, H.~Q. Ngo, M.~Nikolic, D.~Olteanu, and H.~Zhang.
\newblock Counting triangles under updates in worst-case optimal time.
\newblock {\em arXiv preprint arXiv:1804.02780}, 2018.

\bibitem{kara2020trade}
A.~Kara, M.~Nikolic, D.~Olteanu, and H.~Zhang.
\newblock Trade-offs in static and dynamic evaluation of hierarchical queries.
\newblock In {\em Proceedings of the 39th ACM SIGMOD-SIGACT-SIGAI Symposium on
  Principles of Database Systems}, pages 375--392, 2020.

\bibitem{kearns1997teaching}
R.~Kearns, S.~Shead, and A.~Fekete.
\newblock A teaching system for sql.
\newblock In {\em Proceedings of the 2nd Australasian conference on Computer
  science education}, pages 224--231, 1997.

\bibitem{khamis2020functional}
M.~A. Khamis, R.~R. Curtin, B.~Moseley, H.~Q. Ngo, X.~Nguyen, D.~Olteanu, and
  M.~Schleich.
\newblock Functional aggregate queries with additive inequalities.
\newblock {\em ACM Transactions on Database Systems (TODS)}, 45(4):1--41, 2020.

\bibitem{koller2009probabilistic}
D.~Koller and N.~Friedman.
\newblock {\em Probabilistic graphical models: principles and techniques}.
\newblock MIT press, 2009.

\bibitem{kotsis2000elimination}
N.~Kotsis and D.~R. McGregor.
\newblock Elimination of redundant views in multidimensional aggregates.
\newblock In {\em International Conference on Data Warehousing and Knowledge
  Discovery}, pages 146--161. Springer, 2000.

\bibitem{kumar2015learning}
A.~Kumar, J.~Naughton, and J.~M. Patel.
\newblock Learning generalized linear models over normalized data.
\newblock In {\em Proceedings of the 2015 ACM SIGMOD International Conference
  on Management of Data}, pages 1969--1984, 2015.

\bibitem{lakshmanan2008if}
L.~V. Lakshmanan, A.~Russakovsky, and V.~Sashikanth.
\newblock What-if olap queries with changing dimensions.
\newblock In {\em 2008 IEEE 24th International Conference on Data Engineering},
  pages 1334--1336. IEEE, 2008.

\bibitem{lauritzen2003graphical}
S.~L. Lauritzen and N.~A. Sheehan.
\newblock Graphical models for genetic analyses.
\newblock {\em Statistical Science}, pages 489--514, 2003.

\bibitem{leis2015good}
V.~Leis, A.~Gubichev, A.~Mirchev, P.~Boncz, A.~Kemper, and T.~Neumann.
\newblock How good are query optimizers, really?
\newblock {\em Proceedings of the VLDB Endowment}, 9(3):204--215, 2015.

\bibitem{leis2018query}
V.~Leis, B.~Radke, A.~Gubichev, A.~Mirchev, P.~Boncz, A.~Kemper, and
  T.~Neumann.
\newblock Query optimization through the looking glass, and what we found
  running the join order benchmark.
\newblock {\em The VLDB Journal}, 27(5):643--668, 2018.

\bibitem{lepar2013comparison}
V.~Lepar and P.~P. Shenoy.
\newblock A comparison of lauritzen-spiegelhalter, hugin, and shenoy-shafer
  architectures for computing marginals of probability distributions.
\newblock {\em arXiv preprint arXiv:1301.7394}, 2013.

\bibitem{li2005formal}
J.~Li, Z.~A. Talebi, R.~Chirkova, and Y.~Fathi.
\newblock A formal model for the problem of view selection for aggregate
  queries.
\newblock In {\em East European Conference on Advances in Databases and
  Information Systems}, pages 125--138. Springer, 2005.

\bibitem{li2014widetable}
Y.~Li and J.~M. Patel.
\newblock Widetable: An accelerator for analytical data processing.
\newblock {\em Proceedings of the VLDB Endowment}, 7(10):907--918, 2014.

\bibitem{marcus2020bao}
R.~Marcus, P.~Negi, H.~Mao, N.~Tatbul, M.~Alizadeh, and T.~Kraska.
\newblock Bao: Learning to steer query optimizers.
\newblock {\em arXiv preprint arXiv:2004.03814}, 2020.

\bibitem{marcus2019neo}
R.~Marcus, P.~Negi, H.~Mao, C.~Zhang, M.~Alizadeh, T.~Kraska, O.~Papaemmanouil,
  and N.~Tatbul.
\newblock Neo: A learned query optimizer.
\newblock {\em arXiv preprint arXiv:1904.03711}, 2019.

\bibitem{mckinney2010data}
W.~McKinney et~al.
\newblock Data structures for statistical computing in python.
\newblock In {\em Proceedings of the 9th Python in Science Conference}, volume
  445, pages 51--56. Austin, TX, 2010.

\bibitem{miedema2022identifying}
D.~Miedema, E.~Aivaloglou, and G.~Fletcher.
\newblock Identifying sql misconceptions of novices: findings from a
  think-aloud study.
\newblock {\em ACM Inroads}, 13(1):52--65, 2022.

\bibitem{nambiar2006making}
R.~O. Nambiar and M.~Poess.
\newblock The making of tpc-ds.
\newblock In {\em VLDB}, volume~6, pages 1049--1058, 2006.

\bibitem{widetable2}
P.~Navid.
\newblock Modern data warehouse modelling: The definitive guide - part 2.
\newblock \url{https://hightouch.io/blog/data-warehouse-modelling-part-2/},
  2021.

\bibitem{neumann2018adaptive}
T.~Neumann and B.~Radke.
\newblock Adaptive optimization of very large join queries.
\newblock In {\em Proceedings of the 2018 International Conference on
  Management of Data}, pages 677--692, 2018.

\bibitem{ngo2018worst}
H.~Q. Ngo, E.~Porat, C.~R{\'e}, and A.~Rudra.
\newblock Worst-case optimal join algorithms.
\newblock {\em Journal of the ACM (JACM)}, 65(3):1--40, 2018.

\bibitem{nikolic2018incremental}
M.~Nikolic and D.~Olteanu.
\newblock Incremental view maintenance with triple lock factorization benefits.
\newblock In {\em Proceedings of the 2018 International Conference on
  Management of Data}, pages 365--380, 2018.

\bibitem{olteanu2015size}
D.~Olteanu and J.~Z{\'a}vodn{\`y}.
\newblock Size bounds for factorised representations of query results.
\newblock {\em ACM Transactions on Database Systems (TODS)}, 40(1):1--44, 2015.

\bibitem{pearl1982reverend}
J.~Pearl.
\newblock {\em Reverend Bayes on inference engines: A distributed hierarchical
  approach}.
\newblock Cognitive Systems Laboratory, School of Engineering and Applied
  Science~…, 1982.

\bibitem{pineda2015novel}
A.~L. Pineda and V.~Gopalakrishnan.
\newblock Novel application of junction trees to the interpretation of
  epigenetic differences among lung cancer subtypes.
\newblock {\em AMIA Summits on Translational Science Proceedings}, 2015:31,
  2015.

\bibitem{raman2013db2}
V.~Raman, G.~Attaluri, R.~Barber, N.~Chainani, D.~Kalmuk, V.~KulandaiSamy,
  J.~Leenstra, S.~Lightstone, S.~Liu, G.~M. Lohman, et~al.
\newblock Db2 with blu acceleration: So much more than just a column store.
\newblock {\em Proceedings of the VLDB Endowment}, 6(11):1080--1091, 2013.

\bibitem{ramirez2009fault}
J.~C. Ramirez, G.~Munoz, and L.~Gutierrez.
\newblock Fault diagnosis in an industrial process using bayesian networks:
  Application of the junction tree algorithm.
\newblock In {\em 2009 Electronics, Robotics and Automotive Mechanics
  Conference (CERMA)}, pages 301--306. IEEE, 2009.

\bibitem{roy2015explaining}
S.~Roy, L.~Orr, and D.~Suciu.
\newblock Explaining query answers with explanation-ready databases.
\newblock {\em Proceedings of the VLDB Endowment}, 9(4):348--359, 2015.

\bibitem{roy2014formal}
S.~Roy and D.~Suciu.
\newblock A formal approach to finding explanations for database queries.
\newblock In {\em Proceedings of the 2014 ACM SIGMOD international conference
  on Management of data}, pages 1579--1590, 2014.

\bibitem{santos2021correlation}
A.~Santos, A.~Bessa, F.~Chirigati, C.~Musco, and J.~Freire.
\newblock Correlation sketches for approximate join-correlation queries.
\newblock In {\em Proceedings of the 2021 International Conference on
  Management of Data}, pages 1531--1544, 2021.

\bibitem{schleich2021structure}
M.~Schleich.
\newblock Structure-aware machine learning over multi-relational databases.
\newblock In {\em Proceedings of the 2021 International Conference on
  Management of Data}, pages 6--7, 2021.

\bibitem{schleich2019layered}
M.~Schleich, D.~Olteanu, M.~Abo~Khamis, H.~Q. Ngo, and X.~Nguyen.
\newblock A layered aggregate engine for analytics workloads.
\newblock In {\em Proceedings of the 2019 International Conference on
  Management of Data}, pages 1642--1659, 2019.

\bibitem{schleich2016learning}
M.~Schleich, D.~Olteanu, and R.~Ciucanu.
\newblock Learning linear regression models over factorized joins.
\newblock In {\em Proceedings of the 2016 International Conference on
  Management of Data}, pages 3--18, 2016.

\bibitem{shafer1990probability}
G.~R. Shafer and P.~P. Shenoy.
\newblock Probability propagation.
\newblock {\em Annals of mathematics and Artificial Intelligence},
  2(1):327--351, 1990.

\bibitem{shaikhha2020multi}
A.~Shaikhha, M.~Schleich, A.~Ghita, and D.~Olteanu.
\newblock Multi-layer optimizations for end-to-end data analytics.
\newblock In {\em Proceedings of the 18th ACM/IEEE International Symposium on
  Code Generation and Optimization}, pages 145--157, 2020.

\bibitem{staniak2019landscape}
M.~Staniak and P.~Biecek.
\newblock The landscape of r packages for automated exploratory data analysis.
\newblock {\em arXiv preprint arXiv:1904.02101}, 2019.

\bibitem{suri2011counting}
S.~Suri and S.~Vassilvitskii.
\newblock Counting triangles and the curse of the last reducer.
\newblock In {\em Proceedings of the 20th international conference on World
  wide web}, pages 607--614, 2011.

\bibitem{taniar2002parallel}
D.~Taniar and R.~B.-N. Tan.
\newblock Parallel processing of multi-join expansion-aggregate data cube query
  in high performance database systems.
\newblock In {\em Proceedings International Symposium on Parallel
  Architectures, Algorithms and Networks. I-SPAN'02}, pages 51--56. IEEE, 2002.

\bibitem{tao2020computing}
Y.~Tao, X.~He, A.~Machanavajjhala, and S.~Roy.
\newblock Computing local sensitivities of counting queries with joins.
\newblock In {\em Proceedings of the 2020 ACM SIGMOD International Conference
  on Management of Data}, pages 479--494, 2020.

\bibitem{vitter1998data}
J.~S. Vitter, M.~Wang, and B.~Iyer.
\newblock Data cube approximation and histograms via wavelets.
\newblock In {\em Proceedings of the seventh international conference on
  Information and knowledge management}, pages 96--104, 1998.

\bibitem{wang2002condensed}
W.~Wang, J.~Feng, H.~Lu, and J.~X. Yu.
\newblock Condensed cube: An effective approach to reducing data cube size.
\newblock In {\em Proceedings 18th International Conference on Data
  Engineering}, pages 155--165. IEEE, 2002.

\bibitem{wang2013scalable}
Z.~Wang, Y.~Chu, K.-L. Tan, D.~Agrawal, A.~E. Abbadi, and X.~Xu.
\newblock Scalable data cube analysis over big data.
\newblock {\em arXiv preprint arXiv:1311.5663}, 2013.

\bibitem{wu2013scorpion}
E.~Wu and S.~Madden.
\newblock Scorpion: Explaining away outliers in aggregate queries.
\newblock 2013.

\bibitem{xirogiannopoulos2019memory}
K.~Xirogiannopoulos and A.~Deshpande.
\newblock Memory-efficient group-by aggregates over multi-way joins.
\newblock {\em arXiv preprint arXiv:1906.05745}, 2019.

\bibitem{yannakakis1981algorithms}
M.~Yannakakis.
\newblock Algorithms for acyclic database schemes.
\newblock In {\em VLDB}, volume~81, pages 82--94, 1981.

\bibitem{zhou2007lazy}
J.~Zhou, P.-A. Larson, and H.~G. Elmongui.
\newblock Lazy maintenance of materialized views.
\newblock In {\em Proceedings of the 33rd international conference on Very
  large data bases}, pages 231--242, 2007.

\bibitem{zhu2015junction}
F.~Zhu, H.~A. Aziz, X.~Qian, and S.~V. Ukkusuri.
\newblock A junction-tree based learning algorithm to optimize network wide
  traffic control: A coordinated multi-agent framework.
\newblock {\em Transportation Research Part C: Emerging Technologies},
  58:487--501, 2015.

\end{thebibliography}

\clearpage

\appendix


\begin{algorithm}
        \caption{Message Passing and Calibration Algorithm}
        \label{message_pass}
        \begin{algorithmic}[1]
            \State // Pass Message from bag u to v where $u,v \in V$\;
            \Function{PassMessage}{$((E,V), \mathcal{X}, \mathcal{Y})$, u, v}\;
                \State // All the neighbours\;
                \State $N(u) = \{c | c\to u\in\mathcal{E}\}$\;
                \State // All incoming messages from in-neighbours except v\;
                \State $M(u) = \{\mathcal{Y}(i\to u) | i\in N(u)\land i\not\eq v\}$\;
                \State // Compute and store message from u to v\;
                \State $\mathcal{Y}(u\to v) = \sum_{u-v\cap u} \Join \left(M(u) \cup \mathcal{X}^{-1}(u) \right) $\;
            \EndFunction
            \State \;
            \State // Upward Message Passing to root $r \in V$ \;
            \Function{Upward}{$((E,V), \mathcal{X}, \mathcal{Y})$, r}
                \ForAll{Bag $c \in V - r$ from leaves to root r bottom up}\;
                    \State p = parent of c\;
                    \State PassMessage($((E,V), \mathcal{X}, \mathcal{Y})$, c, p)\;
                \EndFor
            \EndFunction
            \State \;
            \State // Downward Message Passing from root $r \in V$ \;
            \Function{Downward}{$((E,V), \mathcal{X}, \mathcal{Y})$, r}
                \ForAll{Bag $p \in V$ from root r to leaves top down}\;
                    \ForAll{child bag c of p}\;
                        \State PassMessage($((E,V), \mathcal{X}, \mathcal{Y})$, p, c)\;
                    \EndFor
                \EndFor
            \EndFunction
            \State \;
            \State // Calibrate Junction Hypertree \;
            \Function{Calibration}{$((E,V), \mathcal{X}, \mathcal{Y})$}
                \State // choose a random bag as root\;
                \State $r \in V $ \;
                \State Upward($((E,V), \mathcal{X}, \mathcal{Y})$, r)\;
                \State Downward($((E,V), \mathcal{X}, \mathcal{Y})$, r)\;
            \EndFunction
            
        \end{algorithmic}
    \end{algorithm}

\begin{algorithm}
        \caption{Join Algorithms}
        \label{alg-join}
        \begin{algorithmic}[1]
        
            \Function{Worst Case Optimal Join implements Join}{$\textbf{R}$}
            \State // Apply worst-case optimal join algorithm\;
            \State \Return $ \Join_{R \in \textbf{R}}$
            \EndFunction
            
            \State
            
            \Function{Indicator Projection implements Join}{$\textbf{R}, \textbf{R}_{db}$}
            \State // Find all attributes in join result\;
            \State $\textbf{U} = \bigcup_{R \in \textbf{R}} S_R$\;
            \State // Find all relations whose schema intersect with $\textbf{U}$ and build indicator projection for them\;
            \State $\textbf{R}_{ind} = \{ \pi^{ind}_{S_R \cap \textbf{U}}(R) |R \in \textbf{R}_{db}\}$
            \State \Return Worst Case Optimal Join($\textbf{R} \cup \textbf{R}_{ind}$)
            \EndFunction

        \end{algorithmic}
    \end{algorithm}

\section{Complexity of Calibrated Junction Hypertree}
\label{complexityAnalysis}
In this section, we study the complexity of \cjt. We start with the background of Fractional Hypertree Width, which is the complexity of for general semiring aggregation query~\cite{abo2016faq,joglekar2015aggregations,olteanu2015size}. The analysis of complexity of \cjt is, however, complicated by three factors: 
\begin{myitemize}
  \item SPJA queries over \cjt will change the relations or the variables to eliminate, which in turn affect the complexity. We use modified \jt to analyze the effect of SPJA queries.
  \item \cjt shared computations by re-using the messages, such that only a sub-hypertree needs to be used for additional messages. 
  \item \cjt additionally supports for dynamic workloads (e.g., updating current relations, and augmenting new relations). These dynamic workloads have a larger complexity as "dangling tuples" can't be removed for future workloads. We use the sub-hypergraph to analyze the effect of danling tuples. 
\end{myitemize}

\subsection{Fractional Hypertree Width} 
We start from Fractional Edge Cover, which takes join graph and the size of each relation as input and outputs the complexity of join result. Quantifying the complexity of final join result is not directly useful because, with Early Marginalization, we can avoid the full join result. We then discuss Fractional Hypertree Width, which outputs the complexity of intermediate join result given the Early Marginalization optimization opportunity.

\stitle{Hypergraph}: Hypergraph is graph $\mathcal{G}$ with vertices $\mathcal{V}$ and hyperedges $\mathcal{E}$, where each hyperedge connects non-empty subset of $\mathcal{V}$. Hypergraph has been widely used to represent join graph, where each vertex represents attribute and hyperedge represents relation.

Given a set of annotated relations $\textbf{R} = \{R_1, R_2 ... , R_n\}$, we assume that attributes to join share the same name and consider natural join $R_1 \Join R_2 ... \Join R_n$. We then build hypergraph $(\mathcal{V}, \mathcal{E})$ for the join, where $\mathcal{V}$ is the set of all attributes $S_{R_1} \cup S_{R_2} ... \cup S_{R_n}$ and $\mathcal{E}$ are the schemas of relations $S_{R_1}, S_{R_2}, ..., S_{R_n}$.

By default, we assume that the set of annotated relations $\textbf{R} = \{R_1, R_2 ... , R_n\}$ will result in a connected hypergraph. It's possible that user wants to compute aggregation queries over relations, where there aren't join keys connecting them, a set of hypergraph will be generated and Cartesian Products have to be computed. Our data structure and applications could be easily extended to support this special case. However, this situation is rare for real-world use case, and we don't discuss this special case here.

\stitle{Fractional Edge Cover.} (aka AGM bound~\cite{atserias2008size}). Given Hypergraph $\mathcal{G} = (\mathcal{E}, \mathcal{V})$ where each hyperedge $e \in \mathcal{E}$ is associated with the size of corresponding relation $|R_e|$, the Fractional Edge Cover number $\rho^*$ is the cost of an optimal solution of the following linear program:
\begin{equation}
\begin{split}
\text{min}& \displaystyle\sum\limits_{E \in \mathcal{E}} log_2(|R_E|) x_E \\
\text{s.t.}& \displaystyle\sum\limits_{E: V \in E}  x_E \geq 1, \forall V \in \mathcal{V}\\
&x_E \geq 0, \forall E \in \mathcal{E}
\end{split}
\end{equation}

Fractional Edge Cover $\rho^*$ has been proved to be tight output size bound of join result $O(|R_1 \Join R_2 ... \Join R_n|) = O(2^{\rho^*})$. 

\begin{example}[Fractional Edge Cover]
Consider the triangle query $\sum_{A,B,C} (R(A,B) \Join S(B,C) \Join T(A,C))$. Suppose that the size of each relation is $O(n)$, Its fractional edge cover number $\rho^* = log_2(n) 1.5$ and the join size of all three relations is bounded by $O(n^{1.5})$.   
\end{example}

\stitle{Worst case optimal join.} Even if the join result is bounded by Fractional Edge Cover, traditional binary join may result in intermediate result asymptomatically larger. Consider the triangle query $\sum_{A,B,C} (R(A,B) \Join S(B,C) \Join T(A,C))$ whose join size is is bounded by $O(n^{1.5})$. However, joining any pair of relations will result in intermediate result with size $O(n^2)$.
To bound the intermediate result size during the execution of join, worst case optimal join~\cite{ngo2018worst} has been proposed to guarantee that the time complexity of evaluating join is proportional to the worst-case output size $O(2^{\rho^*})$. The basic idea of to join multiple relations together and carefully skip tuples impossible to appear in join result.

\stitle{Fractional Hypertree Width.} For semiring aggregation query over join graph, we don't need to fully compute the join result, as we can apply early marginalization to eliminate attributes. Given Junction Hypertree $(\mathcal{T}, \mathcal{X}, \mathcal{Y}, \mathcal{Z})$ of hypergraph $(\mathcal{E}, \mathcal{V})$, we only need to join relations in each bag during Message Passing. Fractional Hypertree Width~\cite{joglekar2015aggregations,abo2016faq} of a given Junction Hypertree is just computing the maximum Fractional Edge Cover placed on bags instead of the whole hypergraph. The fractional hypertree width of Junction Hypertree $fhtw((\mathcal{T}, \mathcal{X}, \mathcal{Y}, \mathcal{Z}))$ is $max_{t\in\mathcal{V}_\mathcal{T}} \rho^*_t$ where $\rho^*_t$ is the Fractional Edge Cover of each bag:
\begin{equation}
\begin{split}
\text{min}& \displaystyle\sum\limits_{E \in \mathcal{E}} log_2(|R_E|)  x_E \\
\text{s.t.}& \displaystyle\sum\limits_{E: V \in E}  x_E \geq 1, \forall V \in \textcolor{blue}{\mathcal{X}(t)}\\
&x_E \geq 0, \forall E \in \mathcal{E}
\end{split}
\end{equation}

Given hypergraph $(\mathcal{E}, \mathcal{V})$, fractional hypertree width of hypergraph $fhtw((\mathcal{E}, \mathcal{V}))$ is the minimum of the fractional hypertree widths over all possible Junction Hypertrees. Finding the minimum fractional hypertree width is known to be NP-hard~\cite{fischl2018general}. 

Notice that, to achieve the time complexity of Fractional Hypertree Width for semiring aggregation, using worst case optimal join alone during message passing is not enough.

\begin{example}[Wost Case Optimal Join insufficient for Fractional Hypertree Width]
Consider the Junction Hypertree in the right of \Cref{fig:count_triangle} for the triangle query $\sum_{A,B,C} (R(A,B) \Join S(B,C) \Join T(A,C))$. While the Fractional Hypertree Width is $O(n^{1.5})$, message from bag ABC to any other bag is $O(n^{2})$.
\end{example}

\stitle{Indicator Projection.}
Indicator projection has been introduced~\cite{abo2016faq} to remove redundant tuples inside message and achieve Fractional Hypertree Width for semiring aggregation. The detailed algorithm of Indicator Projection is in \Cref{alg-join}. Basically, besides relations needed to join for message, we need to further add relations whose schema intersects with join result's to remove redundant tuples. For worst case optimal join, adding more relations to join will make the linear program result smaller as long as their shcema has been covered by other relations.

While Indicator Projection is effective to guarantee that semiring aggregation is bounded by Fractional Hypertree Width, it aggressively eliminates all dangling tuples which might be useful for future queries and reduces the reusability of messages. We introduce Static Calibrated Junction Hypertree, which imposes constraints on the use cases to ensure that Indicator Projection could be safely Indicator Projection while ensure the work sharing opportunities of messages. For use case goes beyond Static Calibrated Junction Hypertree, Indicator Projection is not recommended.

\stitle{Static Calibrated Junction Hypertree.} Calibrated Junction Hypertree is static if the set of relations $\textbf{R} = \{R_1, R_2 ... , R_n\}$ stored is unchanged, tuples inside relations won't be updated, and queries over only a subset of relations is disallowed. Therefore, all queries over Static Calibrated Junction Hypertree are over the join of a fixed set of relations $R_1 \Join R_2 ... \Join R_n$, and dangling tuples eliminated by Indicator Projection will never be used.

Besides limiting work sharing opportunities of messages, Indicator Projection also doesn't improve performance empirically because of the large overhead of projection~\cite{arun2018hypertree}.
Because of these limitations, we will only use Indicator Projection for theoretic analysis and we assume that we use worst case optimal join without Indicator Projection to compute message in this paper unless otherwise specified.

\stitle{Dangling Tuple Free.} We start with Junction Hypertree whose bags are dangling tuple free. This assumption is made to ensure that message passing inside Junction Hypertree is always bounded by its Fractional Hypertree Width even without Indicator Projection, so that we can keep focus on the SPJA annotations and work sharing opportunities inside of Junction Hypertree. We will relax this assumption and discuss general cases later.

 Calibrated Junction Hypertree $(\mathcal{T}, \mathcal{X}, \mathcal{Y}, \mathcal{Z})$ over a set of relations $\textbf{R} = \{R_1, R_2 ... , R_n\}$ is dangling tuple free if it doens't contain empty bag and, for each bag $v \in \mathcal{V}_\mathcal{T}$, the join of relations in v doesn't contain dangling tuples: $\Join_{R \in \mathcal{Y}^{-1}(v)} R = \pi_{\mathcal{X}(v)}(R_1 \Join R_2 ... \Join R_n)$ under 0/1 semiring. When Calibrated Junction Hypertree is dangling tuple free, we don't have to apply Indicator Projection and can still achieve Fractional Hypertree Width for calibration. 

 For the next two sections, we will aasume that the \cjt is dangling tuple free, but relaxes this assumption later.

\subsection{SPJA Query} 

To support SPJA queries over \cjt, we previously use annotations to highlight the changes necessary for the message computations. These changes, however will affect the complexity of the queries:

\stitle{Annotations in SPJA queries.} We consider the effects of all annotations  for a given \jt:
\begin{itemize}
  \item \stitle{$\gamma_A$ :} Group-by will increase the complexity as the attributes can't be early marginalized out. Naturally, if its incoming messages contain additional attributes not contained by this bag, the join size of this bag naturally grows by the product of the domain sizes for attributes. This bound considers the worst case when the cartesian product of group-by attributes has to be computed. However, for group-by attributes from the same relation, their join size could be bounded by the relation size, which could be smaller than the product of domain sizes.  
  \item \stitle{$\sum_A$:} In contrast to $\gamma_A$, canceling out the group-by may decrease the complexity as attributes are early marginalized out.
  \item \stitle{$\overline{R}$:} Exclusion of relations may decrease (e.g., for bag containing two relations whose schema are different and share join key, removing one of them reduces the bag size from $O(n^2)$ to $O(n)$) or increase the complexity (e.g., for bag containing three relations with triangle join, removing one of them increases the bag size from $O(n^{1.5})$ to $O(n^2)$). 
    \item\stitle{$R^*$:} Updating relations will change the size of relation |R|.
  
  \item \stitle{$\sigma_{id}$:} Similar to update, selection will reduce the size of relation given its selectivity. More complex bounds~\cite{joglekar2018s} could be used to study the degree of attributes, which we will leave as a future work.
\end{itemize}
To provide a bound of query over annotated relations, we modify the \jt based on the annotations:  for $\gamma_A$, we add attribute A to all bags, and for $\overline{R}$ we remove the relation R from \jt. We note modified \jt without annoatations achieved the same effect of the original \jt with annotations: attribute A is not marginalized out as A is contains by all bags, and R is removed during join. The SPJA queries are thus bounded by the fhtw of the modified \jt.

We note that the increase in complexity is  {\it not} because \cjt's algorithms have a higher complexity, but because of the different treatments of SPJA queries. For instance, previous works support group-by attributes by creating the a large \jt similar to the modified \jt with a higher fhtw, where the group-by attributes are not eliminated during variable elimination~\cite{abo2016faq}. In this work, we reuse the (potentially smaller) \jt where future group-by attributes could have been eliminated, then increase the complexity to support group-by. If the same \jt whose group-by attributes are not eliminated is used, we achieve the same complexity.

For the rest of the section, we consider queries without annotations for simplicity, but the analysis could be similarly extended to annotated \jt using the modified \jt.
 
\subsection{Fractional Sub-Hypertree Width for work sharing} 

We study the work sharing opportunities of Calibrated Junction Hypertree. Note that we are discussing more general cases than before: in \Cref{cjt_detail} we discuss message re-use between different queries in the same \cjt; here we discuss message re-use between two different queries from potentially different \cjt. 
Given Calibrated Junction Hypertree $\phi^*$, and aggregation query Q over the same semiring, we first consider (not calibrated) Junction Hypertree $\phi$ in which we can execute query Q through message passing to root r. We assume that both $\phi^*$ and $\phi$ are dangling tuple free. Without $\phi^*$, the message passing algorithm takes $ftw(\phi)$, whose bottleneck is the maximum of the fractional edge cover of bags in $\phi$: $max_{V \in \mathcal{V}_\mathcal{T}} \rho^*(\mathcal{Y}^{-1}(v))$.

Now let's consider how we can reuse messages from $\phi^*$ for $\phi$. 
Unlike previous works of junction tree~\cite{koller2009probabilistic} where the underlying probabilistic graphical model is assumed to be fixed, we consider cases when we have different join graphs but share some join relations. At a high level, we want to find part of sub-Hypertrees of $\phi^*$ and $\phi$ that are equivalent, such that their messages are the same and we can directly reuse messages. We start by defining the equivalence of Junction Hypertree, and study when the messages can be reused.

\begin{definition}[Equivalence of bags.]
Bag $u$ and $v$ (potentially from different Junction Hypertree) are equivalent iff $\mathcal{Y}^{-1}(u) = \mathcal{Y}^{-1}(v)$. That is, the set of relations mapped to them are the same.
\end{definition}

We use the equivalence of bags to define isomorphism of Junction Hypertree.

\begin{definition}[Isomorphism of Junction Hypertree]
Given two Junction Hypertrees $(\mathcal{T}_1, \mathcal{X}_1, \mathcal{Y}_1, \mathcal{Z}_1)$ and $(\mathcal{T}_2, \mathcal{X}_2, \mathcal{Y}_2, \mathcal{Z}_2)$, they are isomorphic if there exists a bijection between the vertex $f: \mathcal{V}_{\mathcal{T}_1} \rightarrow \mathcal{V}_{\mathcal{T}_2}$ such that bags are equivalent: $\forall u \in \mathcal{V}_{\mathcal{T}_1}, u = f(u)$, the edges are equivalent: $\forall u, v \in \mathcal{V}_{\mathcal{T}_1}, [u,v] \in \mathcal{E}_{\mathcal{T}_1} \Leftrightarrow [f(u),f(v)] \in \mathcal{E}_{\mathcal{T}_2}$, and the attributes of bags are equivalent $\mathcal{X}_1 = \mathcal{X}_2 \circ f$.
\end{definition}

\begin{definition}[Junction sub-Hypertree.]
Given Junction Hypertree $\phi_1 = (\mathcal{T}_1, \mathcal{X}_1, \mathcal{Y}_1, \mathcal{Z}_1)$ which is symmetric directed and unrooted and $\phi_2 = (\mathcal{T}_2, \mathcal{X}_2, \mathcal{Y}_2, \mathcal{Z}_2)$ with root $r_2 \in \mathcal{V}_{\mathcal{T}_2}$, $\phi_2$ is Junction sub-Hypertree of $\phi_1$ if there exists root in $\phi_1$ such that $\mathcal{T}_2$ rooted in $r_2$ is a subtree of $\mathcal{T}_1$ rooted in $r_1$ ($r_2$ is a node in $r_1$ and $\mathcal{T}_2$ contains all $r_2$'s descendants in $r_1$), and $\mathcal{X}_2, \mathcal{Y}_2, \mathcal{Z}_2$ are equivalent to their counterparts but their domains are restricted to bags and edges in $\mathcal{T}_2$.
\end{definition}

We can use the isomorphism of Junction Hypertree to share the messages between $\phi^*$ and $\phi$:

\begin{prop}
Given Calibrated Junction Hypertree $\phi^*$, and Junction Hypertree $\phi$ which we want to perform message passing to root r, let u and v be bags in $\phi$ where u is the parent of v. Then, the message from v to u could be reused from $\phi^*$ iff: 

(1). the Junction Sub-Hypertree $\phi_{sub}$ of $\phi$ rooted in v is isomorphic to some Junction Sub-Hypertree $\phi^*_{sub}$ of $\phi^*$ rooted in $r^*$. 

(2). the intersection of schema between v and u is equivalent to the intersection of schema between $r^*$ and parent $p^*$ of $r^*$ in $\phi^*$: $\mathcal{X}(v) \cap \mathcal{X}(u) = \mathcal{X}(r^*) \cap \mathcal{X}(p^*)$
\end{prop}

The first condition ensures that the join result is the same, and the second condition ensures that the marginalization attributes are the same. Therefore, the message between u and v is the same as the message between $r^*$ and $p^*$:  $ \mathcal{Z}([v,u]) = \mathcal{Z}([r^*,p^*])$.

\begin{figure}
  \centering
      \includegraphics [scale=0.7] {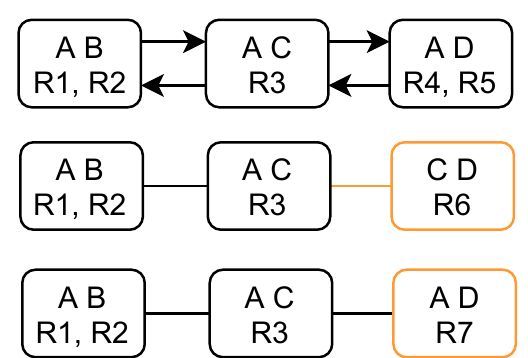}
  \caption{Message Reuse Example}
  \label{fig:msgReuse}
\end{figure}

\begin{example}[Message Reuse]
Consider three Junction Hypertrees in \Cref{fig:msgReuse}, where the first one is calibrated and the rest two are not. For all three Junction Hypertree, the subtrees AB-AC are isomorphic. For the third Junction Hypertree, the message from AC to AD can be reused. For the second Junction Hypertree, the message from AC to CD can't be reused as the second condition is not satisfied (first Junction Hypertree has intersection \{A\}, while second has intersection \{C\}).
\end{example}

\stitle{Fractional Sub-Hypertree Width}
Finally, we study the time complexity of message passing in $\phi$ given the message sharing opportunities. Because of the simplicity of tree structure, we can check which message could be reused from $\phi^*$ efficiently and there are only messages linear to the number of relations to check. We assume that the number of relations is much smaller than the number of rows in relations. Then, the bottleneck of message passing in $\phi$ is the largest bag which can't share computations with $\phi^*$.

Let \texttt{SharedMessages}$(\phi^*, \phi, r)$ be the set of messages from leaves to root r in $\phi$ that could be reused from $\phi^*$ and let \texttt{SubHypertree}$(\phi^*, \phi, r)$ be the set of bags in $\phi$ where there exists some message from this bag not in \texttt{SharedMessages}$(\phi^*, \phi, r)$ plus the root of message passing for absorption. 
The bags in \texttt{SubHypertree}$(\phi^*, \phi, r)$  form a connected tree because if message could be shared between node v to u, so can messages between any descendants of v. We need to perform join and marginalization to bags in \texttt{SubHypertree}$(\phi^*, \phi, r)$ because some of their messages can't reused. The Fractional Sub-Hypertree Width of $\phi$ rooted in r with respect to $\phi^*$ is then the maximum of Fractional Edge Cover placed on these bags: $fhstw_{\phi^*}(\phi, r) = max_{t\in \texttt{SubHypertree}(\phi^*, \phi, r)} \rho^*_t$.

\begin{example}[Fractional Sub-Hypertree Width]
Continue with the example in \Cref{fig:msgReuse}. Given the first Calibrated Junction Hypertree $\phi^*$, for the second Junction Hypertree, its Fractional Sub-Hypertree Width $fhstw_{\phi^*}(\phi, r)$ is the maximum of Fractional Edge Cover placed on bag AC and CD since the message from AC to CD can't be reused and we have to join all relations in AC. For the third Junction Hypertree, its Fractional Sub-Hypertree Width $fhstw_{\phi^*}(\phi, r)$ is only the Fractional Edge Cover placed on bag AD.
\end{example}

We note that $fhstw_{\phi^*}(\phi, r) \leq fhtw_\phi(\phi)$ because \texttt{SubHypertree}$(\phi^*, \phi, r)$ is a subset of bags in $\phi$. The gap between $fhstw_{\phi^*}(\phi, r)$ and $fhtw_\phi(\phi)$ could be as large as the Fractional Edge Cover of bags in $\phi$.

\subsection{Fractional Sub-Hypergraph Sub-Hypertree Width for dynamic workloads} 

We consider the general case where Junction Hypertree is not dangling tuple free and messages between bags are computed without Indicator Projection. We first define Fractional Sub-Hypergraph Hypertree Width, which quantifies the time complexity of calibration for Junction Hypertree without Indicator Projection. Then, we define Fractional Sub-Hypergraph Sub-Hypertree Width, which further exploits the work sharing opportunities.

\stitle{Fractional Sub-Hypergraph Hypertree Width.} Previously, we use Fractional Hypertree Width to study the time complexity of calibration for Junction Hypertree, where Indicator Projection is used to quantify the effect of relations in whole hypergraph. As discussed before, Indicator Projection is not practical and limits the ability of work sharing. Therefore, we consider the time complexity of calibration without Indicator Projection. The main extensions are: (1). consider the join inside a sub-hypergraph instead of whole hypergraph (2). consider the join for each message instead of bag.

Fractional Hypertree Width is defined as the maximum Fractional Edge Cover of bags because join is the most expensive part and all messages sent from each bag has size smaller than the join result (as they are marginalization over join result). However, without marginalization, different messages from the same bag may incur different join sizes, and we need to consider the join separately for different messages. 

Given Calibrated Junction Hypertree $(\mathcal{T}, \mathcal{X}, \mathcal{Y}, \mathcal{Z})$ of hypergraph $(\mathcal{E}, \mathcal{V})$ and directed edge $[v, u] \in \mathcal{E}_\mathcal{T}$, we define \texttt{SubHypergraph}$(\phi, [v, u])$ to be the hypergraph of relations in the bags of subtree $\mathcal{T}_{sub}$ of $\mathcal{T}$ where $\mathcal{T}$ has root u and $\mathcal{T}_{sub}$ has root v. \texttt{SubHypergraph}$(\phi, [v, u])$ is the hypergraph of relations involved with the join when computing messages from v to u. Then, Fractional Sub-Hypergraph Hypertree Width of Junction Hypertree is computing the maximum Fractional Edge Cover of edges over its SubHypergraph: $fsghtw(\mathcal{T}, \mathcal{X}, \mathcal{Y}, \mathcal{Z}) = max_{e \in \mathcal{E}_\mathcal{T}} \rho^*_e$ where $\rho^*_e$ is the Fractional Edge Cover of each bag over $(\mathcal{E}_{sub}, \mathcal{V}_{sub}) = $ \texttt{SubHypergraph}$(\phi, e)$:

\begin{equation}
\begin{split}
\text{min}& \displaystyle\sum\limits_{E \in \textcolor{red}{\mathcal{E}_{sub}}} log_2(|R_E|)  x_E \\
\text{s.t.}& \displaystyle\sum\limits_{E: V \in E}  x_E \geq 1, \forall V \in \textcolor{blue}{\mathcal{X}(u)}\\
&x_E \geq 0, \forall E \in \textcolor{red}{\mathcal{E}_{sub}}
\end{split}
\end{equation}

Given Hypergraph $(\mathcal{E}, \mathcal{V})$, Fractional Sub-Hypergraph Sub-Hypertree Width $fsghtw((\mathcal{E}, \mathcal{V}))$ is the minimum of the Sub-Hypergraph Sub-Hypertree Width over all possible Junction Hypertrees. 

\stitle{Fractional Sub-Hypergraph Sub-Hypertree Width.}  Equipped with Fractional Sub-Hypergraph Hypertree Width, we study the time complexity of message passing in $\phi = (\mathcal{T}, \mathcal{X}, \mathcal{Y}, \mathcal{Z})$ to root r given Calibrated Junction Hypertree $\phi^*$. Let $\mathcal{E}^r_\mathcal{T}$ be a subset of directed edges in $\phi$ when the root is $r \in \mathcal{V}_\mathcal{T}$. The messages we need to compute for message passing to r are: $\mathcal{E}^r_\mathcal{T} - \texttt{SharedMessages}(\phi^*, \phi, r)$. 
Then, the Fractional Sub-Hypergraph Sub-Hypertree Width of $\phi$ rooted in r with respect to $\phi^*$ is the maximum Fractional Edge Cover placed on these bags over sub-hypergraph for all edges e plus the Fractional Edge Cover of root\\  $fsghstw_{\phi^*}(\phi, r) = max(max_{e\in \mathcal{E}^r_\mathcal{T} - \texttt{SharedMessages}(\phi^*, \phi, r)} \rho^*_e, \rho^*_{root})$.

Finally, given a semiring aggregation query Q,  the Fractional Sub-Hypergraph Sub-Hypertree Width of Q with respect to $\phi^*$ is the minimum of the Fractional Sub-Hypergraph Sub-Hypertree Width of all pairs of Junction Hypertree and root capable of executing this query. In general, finding the minimum of Fractional Sub-Hypergraph Sub-Hypertree Width for given query is intractable because there are exponential pairs of Junction Hypertree and root. We can prove that finding the minimum of Fractional Sub-Hypergraph Sub-Hypertree Width is NP-hard: finding the minimum fractional hypertree width is known to be NP-hard~\cite{fischl2018general} and we can reduce this problem to finding the minimum of Fractional Sub-Hypergraph Sub-Hypertree Width by considering an empty $\phi^*$. However, for most applications below where there are interactive queries over the original join graph, we can directly reuse the same Junction Hypertree. When Junction Hypertree and root are fixed, Fractional Sub-Hypergraph Sub-Hypertree Width could be found efficiently.

\section{Optimize Feature Augmentation with \cjt.}
\label{s:augmentation}

While feature augmentations over single join key are efficient, those over multiples multiple join keys are complex. We need to query the \cjt group-by all join keys, which might result in a large Steiner Tree and we need to re-design the \jt after augmentation. 

\stitle{Feature Augmentation over Multiple Bag.} For Feature Augmentation over multiple, we want to query the aggregation group-by the join keys from \cjt. This could be considered as a SPJA query with group-by annotations, and can be computed through Upward Message Passing in the Steiner tree. 

\stitle{Connect Augmentation Relation to \jt.} To connect augmentation relation to \jt where the join key is distributed over multiple bag, we have to add all the join key to the bags of Steiner tree, create an augmentation bag containing augmentation relation, and connect the augmentation bag to any of the bag in Steiner tree. Notice that the \jt with added attributes in the bags can be inefficient, and we may redesign \jt to find a better one.

\stitle{Optimize \cjt design.} To optimize \cjt for feature augmentation, we create empty bags for common join keys. Consider TPC-DS as an example, whose (simplified) join graph (also \jt) is shown in \Cref{fig:tpc_ds_schema}. We can cluster time and stores in an empty bag shown in \Cref{fig:tpc_ds_empty} to support efficient augmentation of spatio-temporal features. 

\section{Minimize Steiner Tree}
\label{s:min_steiner}

In this section, we present the algorithm that, given the \jt with r bags and h group-by annotations, find  the minimum Steiner tree of k annotations in time polynomial to r. We simplify the problem by assuming that each bag has the same size, so the problem is to identify the minimum Steiner tree in terms of the number of bags; our algorithm can be easily extended to the general case.

In \jt, each group-by annotation could be applied to a set of bags containing group-by attributes. We first solve the problem where each annotation could be placed to exactly one bag, then generalizes the algorithm.

\stitle{Single bag annotation.} If each annotation could only be placed on single bag, the problem reduced to: given \jt with r bags and h annotated bags, find the minimum Steiner tree of k annotated bags in terms of the number of bags. We solve the problem by recursion and dynamic programming. We make the edges in \jt bidirectional. For each directed e, it keeps track of x[e][n] defined as "In the sub-tree this edge directs to, what is the minimum number of bags in the Steiner tree that contains the target bag of this edge and n annotated bags (including the target bag if it's annotated)", where n is from 0 to k. This can be computed as follows:

\begin{itemize}
  \item \stitle{Base Case:} For edge e whose target bag is leaf bag, then x[e][1] = 1 and x[e][n] = Inf for n > 1. For all edges, x[e][0] = 0 
  \item \stitle{Recursive Case:} Given edge e and target bag b, let E be the set of edges from b but is not e. One naive way to compute x[e][n] is to consider all possible assignment of the number of annotated bags in to E. This is inefficient as the number of assignments is exponentially large. Instead we combine edges in E one-by-one into one edge. Given two directed edges $e_1$ and $e_2$ in E, we combine them into one directed edge $e^*$ as follows:\\
  $x[e^*][n] = min(x[e_1][m] + x[e_2][n - m]\; for\; m = 0...n)$\\
  Given the final $e^*$ that combines all edges in E, we add the target bag. If the target bag is annotated,  $x[e][n] = x[e^*][n-1] + 1$ for n = 1...k. Otherwise, $x[e][n] = x[e^*][n] + 1$.
\end{itemize}

After we compute the x[e][n] for all directed edges and n from 0 to h, we can find the minimum Steiner tree size by iterating over all edge and compute the minimum Steiner tree containing any end node of this edge. For edge whose two directed edges are $e_1$ and $e_2$, the minimum Steiner tree has size $min(x[e_1][m] + x[e_1][j - m]\; for\; m = 0...n)$.

We analyze the time complexity of the algorithm. We assume both h and k is $O(r)$. For each recursion, each combine takes time $O(r^2)$ (as each $x[e^*][n]$ takes $O(r)$ and there are $O(r)$ n to consider) and there are $O(r)$ combines so $O(r^3)$ in total. There are $O(r)$ directed edges, so $O(r^4)$ to compute all x[e][n]. To find the minimum Steiner tree with x, we iterate $O(r)$ edges and each iteration takes $O(r)$. Therefore, the algorithm takes $O(r^4)$ in total.

\stitle{Multiple bags annotation.} In general, each annotation could be placed on multiple bags. One naive solution is to consider all possible placements, which is exponential in r. Instead, we consider the case where each bag is the root such are all annotations are greedily placed to the bags closest to the root in  $O(r^2)$. There are $O(r)$ possible roots, each could be solved using the previous algorithm in $O(r^4)$, so the final algorithm takes $O(r^5)$.

\begin{figure}[b]
\centering
      \includegraphics [width=0.3\textwidth] {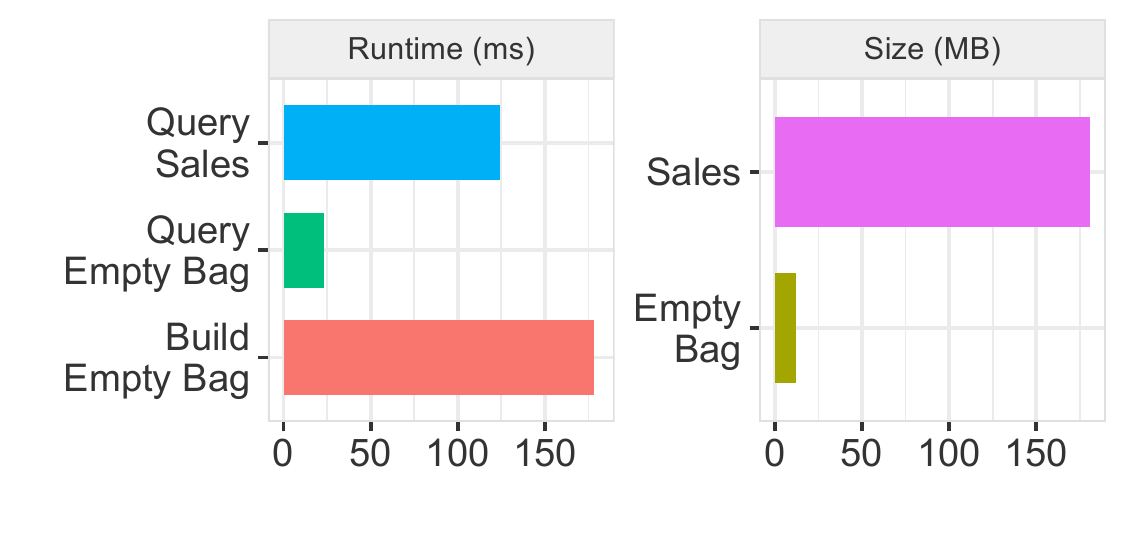}
      \vspace*{-5mm}
      \caption{Runtime to build and query the empty bag and \texttt{Store\_Sales} table, and their storage sizes.}
      \label{exp:empty_bag}
\end{figure} 

\section{Empty Bag Experiment}
Empty bags are a novel \cjt extension to materialize custom views. We evaluate the costs and benefits of empty bag using TPC-DS (SF=1). We create empty bag \texttt{(Store,Time)} as illustrated in \Cref{fig:tpc_ds_empty}. Then, we query the  maximum count of sales for all stores and times: ${\bf Q} = \gamma_{MAX(COUNT)}(\gamma_{COUNT(\cdot), Store, Time}(\Join))$ in two unique ways: (1). Without Empty Bag, ${\bf Q}$ is executed by first aggregating the count over the absorption result of \texttt{Store\_Sales}, since \texttt{Store\_Sales} is the only bag contains both \texttt{Store} and \texttt{Time} dimensions, then computing the max sales. (2). With Empty Bag, ${\bf Q}$ is executed directly over the absorption result of the empty bag, which is sufficient to answer aggregation queries over \texttt{(Store,Time)}.

\Cref{exp:empty_bag} shows the runtimes and sizes. Empty bag takes ${<} 200ms$ to build, while accelerates ${\bf Q}$ by ${\sim}8\times$. The space overhead of empty bag is $15\times$ smaller than \texttt{Store\_Sales}.
\section{Optimize latency with \cjt}
\label{latency_cjt}

\begin{figure}
  \centering
    \begin{subfigure}[b]{0.2\textwidth}
         \centering
         \includegraphics[width=0.7\textwidth]{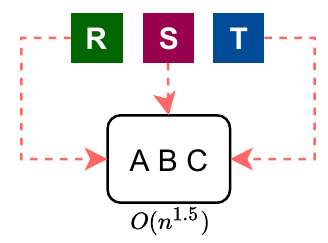}
         \caption{Non-redundant design.}
         \label{fig:count_triangle_reduced}
     \end{subfigure}
    \hfill
     \begin{subfigure}[b]{0.2\textwidth}
         \centering
         \includegraphics[width=\textwidth]{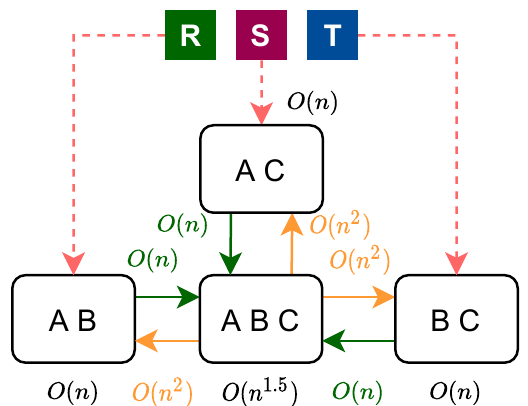}
         \caption{Redundant design.}
         \label{fig:count_triangle_redundant}
     \end{subfigure}
     \hfill
  \caption{Different design for count triangle. }
  
\end{figure}

\stitle{Optimize latency with \cjt.} 
F-IVM provides a trade-off between the size of \cjt and update latency for cyclic join graph. The key insight is that, when there are multiple relations mapped to the same bag, we can create a new bag for individual relation and connect them with an empty bag, which might increase space but reduces latency. This trade-off is critical for Stream Query Processing, where latency is an important metric~\cite{abadi2005design}. Consider the counting triangles under update~\cite{alon1997finding,eden2017approximately, kara2018counting}:

\begin{example}[Count Triangle Under Update]
Given the triangle relations R(A,B), S(B,C), T(A,C) each with size $O(n)$, we want to count the total triangle count under update (one tuple is added to one relation). The \cjt with the minimum size is shown in \Cref{fig:count_triangle_reduced}, where the bag size is $O(n^{1.5})$. We call this \cjt the non-redundant design. The alternative design is to place relations in different bags and connect them with an empty bag as shown in \Cref{fig:count_triangle_redundant}. We call this \cjt the redundant design, which has a larger time and space complexity as the messages from the empty bag are $O(n^2)$.

While redundant design has a larger space complexity, it has a smaller update latency. \Cref{table:count_design} demonstrates the trade-off. Redundant design has an initial overhead of $O(n^2)$ to build \cjt, but after that, each update takes $O(1)$ latency to see the updated query result and the $O(n)$ calibration could take place in the background. On the contrary, for non-redundant design, calibration and update query result happens together, and users have to wait $O(n)$ for the updated result.
\end{example}

\begin{table}[h!]
\begin{center}
\setlength{\tabcolsep}{0.2em} 
\begin{tabular}{ m{4.5em}  m{3em}  m{4em}  m{3em} m{10em} } 
 & \textbf{N} & \textbf{N-IVM} & \textbf{R}  & \textbf{R-IVM}\\ 
Space  & $n^{1.5}$ & $n^{1.5}$ & $n^2$ & $n^2$ \\ 
Latency & $n^{1.5}$ & $n$  & $1$  & $1$\\ 
Calibration & $n$  & $n$ & $n^2$  &$n^2$ \gray{initial}, $n$ \gray{update}\\ 
\end{tabular}

\end{center}
\caption{Design trade-off. R stands for redundant and N stands for non-redundant.}
\label{table:count_design}
\end{table}

While previous works~\cite{abo2016faq,schleich2019layered,nikolic2018incremental} focus on non-redundant design to optimize single query, the combination of \cjt and F-IVM presents a novel trade-off. As a rule of thumb, when users are interested in the live streaming result, we should place frequently updated relations in different bags for low latency. 

\begin{table}[h!]
\begin{center}
\setlength{\tabcolsep}{0.2em} 
\begin{tabular}{ | m{1em} | m{13.5em} | m{4.5em} | m{4.5em}|} 
\hline
 & Task & Reduced & Redundant \\ 
\hline
\multirow{4}{1em}{B} & Calibration + \(\gamma_{COUNT(\cdot)}(\Join)\) & 126.18 & 621.65 \\ 
 &  Update BC (1) & 5.27 & 0.02 \\ 
 &  Update BC (100) & 6.45 & 0.05 \\ 
 &  Augment BC (10000) & 154.66 & 6.41 \\ 
\hline
\multirow{4}{1em}{U} & Calibration + \(\gamma_{COUNT(\cdot)}(\Join)\) & 11.24 & 28332.29 \\ 
 &  Update BC (1) & 5.28 & 0.01 \\ 
 &  Update BC (100) & 6.14 & 0.05 \\ 
 &  Augment BC (10000) & 13.56 & 6.57 \\ 
\hline

\end{tabular}

\end{center}
\caption{Dataset result in millisecond. B for balanced. U for unbalanced.}
\label{table:design_trade_off}
\end{table}

\stitle{Experiment setup.}
We compare two Junction Hypertree designs for counting triangle problem \(\gamma_{COUNT}(R[A,B]\Join S[B,C]\Join T[A,C])\) on local C++ implemented query engine. There are two designs for cyclic join graph. The first design \texttt{Reduced} uses one unified tree node for all relations to retain the cycle. The second design \texttt{Redundant} breaks the cycle using one tree node for each relation, which are connected by an empty node in the middle. These two designs have different complexity for different tasks. 

We consider two synthetic datasets: Balanced vs unbalanced. For balanced workload, each relation is cartesian product \([100] \times [100]\), where n is a unary relation with tuples from 1 to n. Each relation has size 10000 and the join size $10000^{1.5}$ which reaches the worst case bound. For unbalanced dataset, we generates \([1] * [10000]\) for AB and AC where A is highly biased, and \([100] \times [100]\) for BC. The join result is then only  $10000$. However, the message from empty tree node ABC to BC is $10000^2$. We consider three tasks: Calibration for count, update count when 1 or 100 tuples in BC is updated, and augement BC. We choose BC is because BC is same in two datasets, and we want to fix the relation to study how other datasets affect. The task over other tables will scale proportional to their sizes.

\stitle{Takeaways.} The experiment result is in \Cref{table:design_trade_off}.
For calibration, we find that unbalanced dataset is much larger for redundant, which is expected as it has larger theoretic bound. The calibration time for reduced is much smaller because small join size as expected. However, redundant design significantly reduces update latency by $>100\times$. If calibration could be computed offline, it is worthwhile to place relation that is frequently updated and queried into a single bag. 
\section{Exploratory Query}
\label{exp:local_full}

We first study the performance improvement of queries over calibrated junction hypertree. 

\stitle{Datasets.} We concisder four datasets: (1). Favorita~\cite{favorita} is a public dataset for purchasing and sales forecasting. Favorita has also been used by LMFAO ~\cite{schleich2019layered} for factorized learning. (2). Lego~\cite{lego} is a public dataset for the inventories of every official LEGO set. (3). IMDB~\cite{imdb} is a public dataset containing information about movies and related facts about actors, production companies, etc. IMDB has been widely used as a benchmark for query optimizer~\cite{marcus2020bao,leis2015good,marcus2019neo}. (4). TPC-DS~\cite{nambiar2006making} (scale factor 1) is a synthetic dataset for decision support benchmark with star schema. 

Similar to other works~\cite{aberger2017emptyheaded,raman2013db2}, we apply dictionary encoding as preprocess to map original data values to 32-bit unsigned integers. Dictionary encoding eliminates fields with large strings and simplifies the operator design.

\stitle{Competitors.} We consider two competitors to Calibrated Junction Tree: (1). LMFAO~\cite{schleich2019layered} which is the state-of-art system that use junction tree to optimize for semi-ring aggregation queries. LMFAO generate efficient codes to execute queries, but it destroys all the intermediate results after query execution and can't apply IVM to maintain them. (2). Junction Hypertree-IVM is a variant of Calibrated Junction Hypertree, which doesn't calibrate the junction hypertree and, for each query, it passes message across the whole hypertee. However, it applies factorisied incremental view maintainance~\cite{nikolic2018incremental} to pass delta messages for what-if queries. Similar to LMFAO, we need to choose root for the junction tree to which all messages converge. For the experiment, we try all possible roots and report the one with the smallest run time.

\stitle{Workloads.}  For all queries, we use COUNT as the semi-ring aggregation for simplicity. More complex aggregation has a large constant overhead, which won't affect the relative performance. 
Each query involves dimension tables, and we choose dimension tables popular in related queries for IMDB and TPC-DS, or in Kaggle notebook for Favorita and Lego. We consider three types of queries as workloads for experiment: (1). OLAP queries: we query the total count of join result, and add group by and dummy selection (selection doesn't exclude any tuple but force message passing) over the primary keys of dimension tables. (2). Data Explanation: we query the total count of join result after removing a small number of tuples in dimension tables. We choose 10 as the number of tuples to remove, and report the percentage of 10 tuples in the original dimension table. (3). Augmentation: we choose dimension table for augmentation and query the total count after augmentation. We generate an augmented table which maps primary key to random generated numbers.

\stitle{Takeaways.} The experiment result is shown in \Cref{exp:exploratoryquery}. Calibrated Junction Hypertree has the overhead of calibration, but it's orders of magnitudes faster for all three workloads by aggressively reusing the messages. We note that the calibration overhead of TPC-DS dataset is large compared ($\times20$ compared to single query in LMFAO). This is because TPC-DS has star schema, where store\_sales, the large fact table, is connected with many dimension tables, and all messages from store\_sales are expensive. However, the expensive cost of calibration is one-time, and IVM could be applied to maintain the calibration for future updates to fact tables, as analyzed in the next section.

LMFAO outperforms Junction Hypertree-IVM for most OLAP and Augementation workloads. This is because of the implementation difference: We implement hash-based worst case optimal join~\cite{freitag2020adopting}, which has the overhead of iterating through trie, while LMFAO uses naive sort-merge join. LMFAO requires all the cyclic join to be pre-computed and only consider acyclic join graph as junction hypertree. LMFAO doesn't work for the common count and maintain triangles problem~\cite{suri2011counting}. Besides, LMFAO is heavily optimized with efficient code generations.

Even if LFMAO is heavily optimized, Junction Hypertree-IVM outperforms Lego for some OLAP and Augmentation queries. This is mainly because LMFAO makes poor root choice in junction hypertree. The Inventory relation in Lego dataset is sparse, which causes the messages size asymmetric. LMFAO chooses root based on simple heuristics, which doesn't consider the sparsity of relation. 

Finally, for what-if query, factorized IVM has a significant performance improvement especially when the ratio of tuples in original relation is small and calibration further improves the performance on top of factorized IVM. While factorized IVM allows the messages to be smaller, calibration further avoids unnecessary passing of messages.

\begin{table}[h!]
\begin{center}
\setlength{\tabcolsep}{0.2em} 
\begin{tabular}{ | m{1em} | m{13.5em} | m{3.5em} | m{3.5em}| m{3.5em} |} 
\hline
 & Task & LMFAO & \jt-IVM  & \cjt \\ 
\hline
\multirow{9}{1em}{F} & Calibration &   &  & 2108.32  \\ 
 & \(\gamma_{COUNT(\cdot)}(\Join)\) & 179.88 & 943.12 & 0.07\\ 
 & \(\gamma_{COUNT(\cdot), Store}(\Join)\) & 180.20 & 1151.70 & 0.06\\ 
 & \(\gamma_{COUNT(\cdot), Oil} \sigma_{Store}(\Join)\) & 179.71 & 1211.48 & 1.54\\ 
 & Remove 10 Items (0.24\%)  & 176.25 & 2.76 & 0.05\\ 
 & Remove 10 Stores (18.86\%)  & 177.33 & 24.21 & 0.03\\ 
 & Remove 10 Transactions (0.01\%)  & 179.24 & 5.63 & 0.02\\ 
 & Augment Items (4100)  & 181.80  & 973.51 & 3.12\\ 
 & Augment Holiday (1734)  & 180.74  & 1197.44 & 0.27\\ 
\hline
\multirow{7}{1em}{L} & Calibration &   &  & 393.64 \\ 
 & \(\gamma_{COUNT(\cdot)}(\Join)\) & 54.15 & 13.84 & 0.09\\ 
 & \(\gamma_{COUNT(\cdot), Set}(\Join)\) & 53.31 & 369.42 & 0.15\\ 
 & \(\gamma_{COUNT(\cdot), Part} \sigma_{97.10\% Color}(\Join)\) & 56.39 & 14.38 & 0.65\\ 
 & Remove 10 Color (7.46\%)  & 54.61 & 127.75 & 0.02\\ 
 & Remove 10 Themes (1.63\%)  & 52.15 & 5.47 & 0.03\\ 
 & Augment Color (135)  & 55.91 & 30.74 & 0.09\\ 
\hline
\multirow{8}{1em}{I} & Calibration &   &  & 43796.63  \\ 
 & \(\gamma_{COUNT(\cdot)}(\Join)\) & 14722.10 & 18273.66 & 0.05\\ 
 & \(\gamma_{COUNT(\cdot), Person}(\Join)\) & 14854.87 & 24173.64 & 552.02\\ 
 & \(\gamma_{COUNT(\cdot), Movie} \sigma_{Company}(\Join)\) & 13599.41 & 17039.82 & 682.48 \\ 
 & Remove 10 Person (0.0002 \%)  & 14753.22 & 18.11 & 0.12\\ 
 & Remove 10 Movie-key (0.007\%)  & 14833.76 & 2315.08 & 0.15\\ 
 & Augment Person (4167491)  & 14953.03  & 24932.62 & 932.25\\ 
 & Augment Company (234997)  & 14784.58  & 18567.45 & 275.66\\ 
\hline
\multirow{8}{1em}{T} & Calibration &   &  & 9402.95  \\ 
 & \(\gamma_{COUNT(\cdot)}(\Join)\) & 482.53 & 653.70 & 0.06\\ 
 & \(\gamma_{COUNT(\cdot), Store}(\Join)\) & 483.13 & 1737.36 & 0.12\\ 
 & \(\gamma_{COUNT(\cdot), Year} \sigma_{Store}(\Join)\) & 478.75 & 1236.53 & 638.61\\ 
 & Remove 10 Item (0.06\%)  & 480.83 & 22.01 & 0.15\\ 
 & Remove 10 Customer (0.1\%)  & 484.98 & 18.25 & 0.05\\ 
 & Augment Date (201)  & 484.61  & 652.39 & 0.12\\ 
 & Augment Address (1920800)  & 498.53  & 1931.30 & 305.53\\ 
\hline
\end{tabular}

\end{center}
\caption{Exploratory query. In millisecond.}
\label{exp:exploratoryquery}
\end{table}
\section{TPC-H details}
\label{s:tpchqueries}
We discussed how we rewrite TPC-H queries into semi-ring SPJA queries.

\stitle{Query 3.} We remove top and order-by. We also remove $L_ORDERKEY$ group-by because otherwise the result has too many groups.

\begin{verbatim}
SELECT	SUM(L_EXTENDEDPRICE*(1-L_DISCOUNT))	AS REVENUE,
O_ORDERDATE, O_SHIPPRIORITY
FROM	CUSTOMER, ORDERS, LINEITEM
WHERE	C_MKTSEGMENT	= 'FURNITURE' AND
C_CUSTKEY	= O_CUSTKEY AND L_ORDERKEY	= O_ORDERKEY AND
O_ORDERDATE	< '1995-03-28' AND L_SHIPDATE	> '1995-03-28'
GROUP	BY	O_ORDERDATE, O_SHIPPRIORITY;
\end{verbatim}

\stitle{Query 4.}
We rewrite the nested query and remove order-by and distinct.

\begin{verbatim}
SELECT	O_ORDERPRIORITY, count(distinct  O_ORDERKEY)
FROM	LINEITEM, ORDERS
WHERE	O_ORDERDATE	>= '1997-04-01' AND
	O_ORDERDATE	< cast (date '1997-04-01' + interval '3 months' as date) 
AND L_ORDERKEY	= O_ORDERKEY AND L_COMMITDATE	< L_RECEIPTDATE
GROUP	BY	O_ORDERPRIORITY;
\end{verbatim}

\stitle{Query 5.}
For query 5, we break cycle with additional optimization.

\begin{verbatim}
SELECT	N_NATIONKEY,
	SUM(L_EXTENDEDPRICE*(1-L_DISCOUNT))	AS REVENUE
FROM	CUSTOMER, ORDERS, LINEITEM, SUPPLIER, NATION, REGION
WHERE	C_CUSTKEY	= O_CUSTKEY AND	L_ORDERKEY	= O_ORDERKEY AND
	L_SUPPKEY	= S_SUPPKEY AND	C_NATIONKEY	= S_NATIONKEY AND
	S_NATIONKEY	= N_NATIONKEY AND	N_REGIONKEY	= R_REGIONKEY AND
	R_NAME		= 'MIDDLE EAST' AND  o_orderdate >= date '1994-01-01' AND 
o_orderdate < cast (date '1994-01-01' + interval '1 year' as date)
GROUP	BY	N_NATIONKEY;
\end{verbatim}

\stitle{Break cycle for $Q_5$.} $Q_5$ joins customer and supplier by nation, which makes the join graph cyclic and \jt expensive. Luckily, $Q_5$ also group-by nation. We discuss the technique to break cycle: rewrite join + group-by as a set of selections.

Consider the cyclic join of R(A,B), S(A,C), T(B,C). If we know that all future queries will group-by attribute A, we can break the cycle through query rewriting. The original join query is:

\indent\texttt{SELECT A, COUNT(*)}\\
\indent\texttt{FROM R(A,B), S(A,C), T(B,C)}\\
\indent\texttt{WHERE R.A = S.A AND R.B = T.B AND S.C = T.C}\\
\indent\texttt{GROUP BY R.A}

The group by query could be considered as a set of smaller queries, each select a value of A in its domain dom(A). Therefore, for each $a\in dom(A)$, we query

\indent\texttt{SELECT COUNT(*)}\\
\indent\texttt{FROM R(A,B), S(A,C), T(B,C)}\\
\indent\texttt{WHERE \text{\sout{R.A = S.A AND}} R.B = T.B AND S.C = T.C \textcolor{blue}{AND R.A = a AND S.A = a}}

The rewritten query has acyclic join graph. This optimization is closely related to conditioning in Probabilistic graphical model~\cite{koller2009probabilistic}.

\stitle{Query 7.}
We rewrite the nested query and remove order-by.

\begin{verbatim}
SELECT	N1.N_NAME			AS SUPP_NATION,
N2.N_NAME			AS CUST_NATION,
extract(year from L_SHIPDATE) as L_YEAR,
SUM(L_EXTENDEDPRICE*(1-L_DISCOUNT))	AS VOLUME
FROM	SUPPLIER, LINEITEM, ORDERS, CUSTOMER, NATION N1, NATION N2
WHERE	S_SUPPKEY	= L_SUPPKEY AND
O_ORDERKEY	= L_ORDERKEY AND C_CUSTKEY	= O_CUSTKEY AND
S_NATIONKEY	= N1.N_NATIONKEY AND C_NATIONKEY	= N2.N_NATIONKEY AND
(	(N1.N_NAME	= 'UNITED STATES'	AND N2.N_NAME	= 'JAPAN') OR
(N1.N_NAME	= 'JAPAN'	AND N2.N_NAME	= 'UNITED STATES')
) AND L_SHIPDATE	BETWEEN '1995-01-01' AND '1996-12-31'
GROUP	BY	SUPP_NATION, CUST_NATION, L_YEAR
\end{verbatim}

\stitle{Query 8.}
We only consider the inner query, as outer query is cheap to compute.

\begin{verbatim}
SELECT	extract(year from o_orderdate) as o_year, 
SUM(L_EXTENDEDPRICE * (1-L_DISCOUNT)), N2.N_NATIONKEY
FROM	PART, SUPPLIER, LINEITEM, ORDERS, 
CUSTOMER, NATION N1, NATION N2, REGION
WHERE	P_PARTKEY	= L_PARTKEY AND S_SUPPKEY	= L_SUPPKEY AND
		L_ORDERKEY	= O_ORDERKEY AND O_CUSTKEY	= C_CUSTKEY AND
		C_NATIONKEY	= N1.N_NATIONKEY AND N1.N_REGIONKEY	= R_REGIONKEY AND 
		R_NAME		= 'ASIA' AND S_NATIONKEY	= N2.N_NATIONKEY AND
		O_ORDERDATE	BETWEEN '1995-01-01' AND '1996-12-31' AND
		P_TYPE		= 'MEDIUM ANODIZED COPPER'
GROUP BY N2.N_NATIONKEY, o_year;
\end{verbatim}

\section{Space Overhead of Data Cube}
\begin{figure}
  \includegraphics[width=0.8\columnwidth]{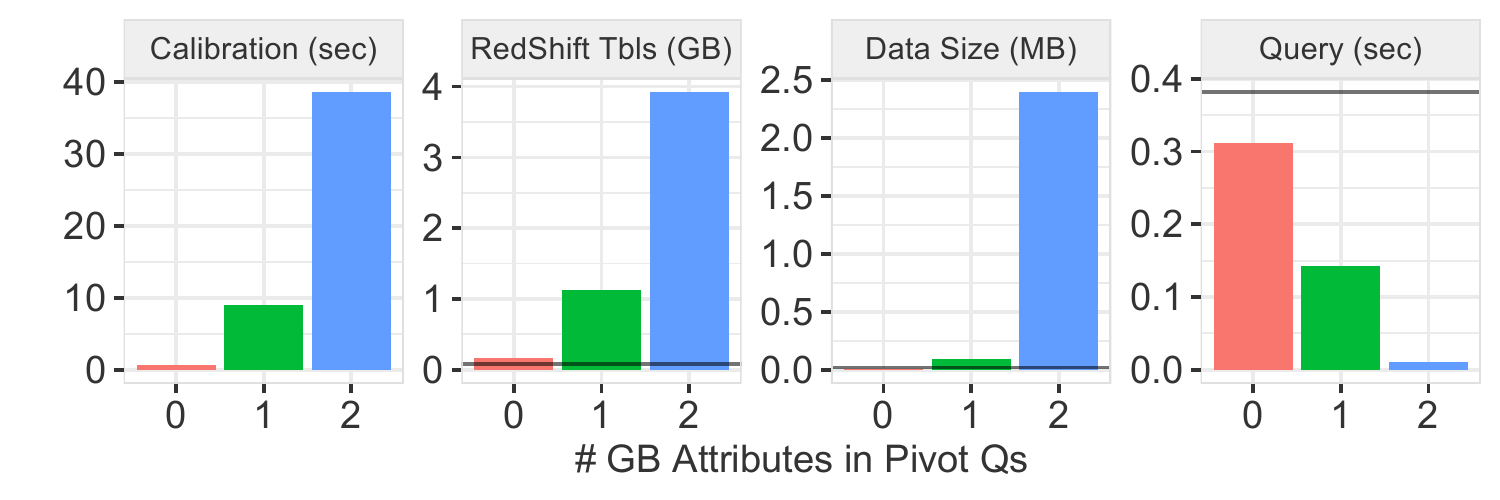}
  \caption{
   RedShift pads tables to be at least $6MB$, which penalizes many small tables.   So we report the total RedShift table size, and actual data size.   Horizontal lines represent, from left-to-right: base DB table size, base DB actual data size, and average runtime with \jt
  }  
\end{figure}

The total Redshift table sizes created during calibration is exponential in $k$, and consistent with the analysis in \Cref{app:cube_complexity}.  Redshift appears to pad small tables to ${>}6MB$, hence the large sizes (${\sim}4 GB$ for $k=2$). Unfortunately. the tables cannot be naively compacted (by unioning into a single table) because their schemas are different. Thus we report the actual {\it Data Size} by adding tuple size times cardinality across the tables. The overhead is only $0.17\times$ for $k=0$, $5\times$ for $k=1$ and $127.73\times$ for $k=2$. Given the significant query performance improvement, the space-time trade-off may be worthwhile.

\end{document}